\documentclass[10pt,journal,romanappendices]{IEEEtran}
\usepackage{stmaryrd}
\usepackage{mathrsfs}
\usepackage[english]{babel}
\usepackage{color}
\usepackage[final]{graphicx}
\usepackage{epstopdf}
\usepackage{multirow}
\usepackage{amsmath}
\usepackage{amssymb,times}
\usepackage{cite}
\usepackage{algorithm}
\usepackage{algorithmic}
\usepackage{stfloats}
\usepackage{subfigure}
\usepackage{wasysym}
\usepackage{hyperref}
\usepackage{cleveref}

\newcommand{\ba}{\begin{array}}
\newcommand{\ea}{\end{array}}
\newcommand{\be}{\begin{displaymath}}
\newcommand{\ee}{\end{displaymath}}
\newcommand{\ben}{\begin{equation}}
\newcommand{\een}{\end{equation}}
\newcommand{\bena}{\begin{eqnarray}}
\newcommand{\eena}{\end{eqnarray}}
\newcommand{\beqa}{\begin{eqnarray*}}
\newcommand{\enqa}{\end{eqnarray*}}

\newcommand{\bc}{\begin{center}}
\newcommand{\ec}{\end{center}}
\newcommand{\bi}{\begin{itemize}}
\newcommand{\ei}{\end{itemize}}
\newcommand{\benu}{\begin{enumerate}}
\newcommand{\eenu}{\end{enumerate}}
\newcommand{\bdes}{\begin{description}}
\newcommand{\edes}{\end{description}}
\newcommand{\bt}{\begin{tabular}}
\newcommand{\et}{\end{tabular}}

\newcommand \betabf{\boldsymbol{\beta}}

\newcommand \varepsilonbf{\boldsymbol{\varepsilon}}

\newcommand \mubf{\boldsymbol{\mu}}
\newcommand \nubf{\boldsymbol{\nu}}
\newcommand \xibf{\boldsymbol{\xi}}

\newcommand \varphibf{\boldsymbol{\varphi}}

\newcommand \Sigmabf{\boldsymbol{\Sigma}}

\newcommand \hbf{{\bf h}}

\newcommand \sbf{{\bf s}}

\newcommand \ubf{{\bf u}}

\newcommand \wbf{{\bf w}}
\newcommand \xbf{{\bf x}}
\newcommand \ybf{{\bf y}}

\newcommand \Cbf{{\bf C}}

\newcommand \Ibf{{\bf I}}

\newcommand \Sbf{{\bf S}}

\newcommand \Xbf{{\bf X}}
\newcommand \Ybf{{\bf Y}}

\newcommand{\Cset}{{\mathbb C}}

\newcommand{\Hcal}{\mathcal{H}}

\newcommand{\Xcalbf}{\boldsymbol{\mathcal{X}}}

\newcommand{\gs}{\mathop{\gtrless}\limits}

\newcommand{\circlambda}{\mbox{$\Lambda$
             \kern-.85em\raise1.5ex
             \hbox{$\scriptstyle{\circ}$}}\,}

\newtheorem{Theorem}{Theorem}
\newcommand{\ThmRef}[1]{\ref{thm:#1}}
\newcommand{\ThmLabel}[1]{\label{thm:#1}}

\IEEEoverridecommandlockouts

\begin{document}

\title{Signal Detection in Distributed MIMO Radar with Non-Orthogonal Waveforms and Sync Errors
}
\author{Hongbin Li, 
Fangzhou Wang, 
Cengcang Zeng,
 and Mark A. Govoni
\thanks{H. Li, F. Wang, and C. Zeng are with the Department of Electrical and Computer Engineering, Stevens Institute of Technology, Hoboken, NJ 07030 USA (e-mail: hli@stevens.edu; fwang11@stevens.edu; czeng2@stevens.edu).}
\thanks{M. A. Govoni is with Army Research Laboratory, Aberdeen Proving Ground, MD 21005 USA.}
}

\maketitle

\begin{abstract}
Although routinely utilized in literature, orthogonal waveforms may
lose orthogonality in distributed multi-input multi-output (MIMO)
radar with spatially separated transmit (TX) and receive (RX)
antennas, as the waveforms may experience distinct delays and Doppler
frequency offsets unique to different TX-RX propagation paths. In such
cases, the output of each waveform-specific matched filter (MF),
employed to unravel the waveforms at the RXs, contains both an
\emph{auto term} and multiple \emph{cross terms}, i.e., the filtered
response of the desired and, respectively, undesired waveforms. We
consider the impact of non-orthogonal waveforms and their cross terms
on target detection with or without timing, frequency, and phase
errors. To this end, we present a general signal model for distributed
MIMO radar, examine target detection using existing
coherent/non-coherent detectors and two new detectors, including a
hybrid detector that requires phase coherence locally but not across
distributed antennas, and provide a statistical analysis leading to
closed-form expressions of false alarm and detection probabilities for
all detectors. Our results show that cross terms can behave like
\emph{foes} or \emph{allies}, respectively,
if they and the auto term add \emph{destructively} or
\emph{constructively}, depending on the propagation delay, frequency,
and phase offsets. Regarding sync errors, we show
that phase errors affect only coherent detectors, frequency errors
degrade all but the non-coherent detector, while all are impacted by
timing errors, which result in a loss in the signal-to-noise ratio
(SNR).
\end{abstract}


\section{Introduction}
\label{sec:introduction}
\subsection{Background}
Multi-input multi-output (MIMO) radar, equipped with multiple transmit/receive (TX/RX) antennas, has been of significant interest for civilian and military applications in recent years \cite{LiStoica2007,HaimovichBlum08,WangLiHimedNon11,Govoni2015,AubryMaioHuang16,ChengHeLiaoFang18,
CuiMonga2020,MaioVenturinoISIT08,HeBlumTSP11,GodrichPoor2011,
HeBlumHaimovich10,LiangLeungSo2016,HeLehmannBlum10,WangLiParametric13,YiBlum2020,AbramovichFrazer08,
AkcakayaNehorai11,WangLi2013Radarconf,WangLi2020,SunTangJunTSP14,YangBlum11,AkcakayaNehoraiPhase10,
HeBlumCRB10,DingChen2015,ZhangVorobyov2016,HuLiuLiu2016,WangLi20RadarConf}. There are two broad categories, namely \emph{co-located MIMO radar} (e.g., \cite{LiStoica2007}), where the antennas in the TX and, respectively, RX array are closely spaced (to within a few wavelengths), and \emph{distributed MIMO radar} \cite{HaimovichBlum08}, where the antennas are widely separated from each other. A distributed MIMO radar can be deployed with its sensors placed close to the radar scene (e.g., via unmanned aerial vehicles) and probe the scene from different aspect angles, allowing the radar to exploit the spatial and geometric diversity to enhance target detection and localization performance \cite{HaimovichBlum08,WangLiHimedNon11,MaioVenturinoISIT08,HeBlumTSP11}. A large body of works have been devoted to the development of distributed MIMO radar related techniques, such as power allocation \cite{GodrichPoor2011}, antenna placement \cite{HeBlumHaimovich10}, neural network based optimization \cite{LiangLeungSo2016}, detection in homogeneous \cite{HeLehmannBlum10} and non-homogeneous \cite{WangLiParametric13} clutter environments, low-complexity multi-target detection and localization \cite{YiBlum2020}, among others.

This paper considers target detection in distributed MIMO radar with non-orthogonal waveforms. A MIMO radar transmits multiple waveforms from its TXs to probe the environment. The RXs employ a set of matched filters (MFs), one for each waveform, which are intended to unravel the radar echoes and separate the information carried by different waveforms. Under the condition that the waveforms are orthogonal with zero cross-correlation, a frequently used assumption in the literature, the MF output would contain only the filtered echo of the desired waveform, i.e., the \emph{auto term}, but no \emph{cross terms} from the undesired waveforms, thus resulting in ideal waveform separation. However, it is impossible to maintain orthogonality with arbitrary delays and frequency shifts \cite{AbramovichFrazer08}. The problem becomes more severe in distributed MIMO radar, since the received signals are inherently \emph{asynchronous}, i.e., waveforms sent from different TXs in general cannot simultaneously arrive at an RX due to different propagation distances. Moreover, different TX-RX pairs observe distinct Doppler frequencies for the same moving target due to different aspect angles \cite{WangLiHimedNon11}. Such delay and frequency offsets would render orthogonal waveform on transmit non-orthogonal at the RXs. Hence, the often neglected cross terms can be significant.

The effects of cross terms were examined in \cite{AkcakayaNehorai11}, which treat them as deterministic unknowns, whereas in \cite{WangLi2013Radarconf,WangLi2020}, they were modeled as random quantities with an unknown covariance matrix. In either case, the waveform correlation, which is known, was not utilized. We introduce herein an efficient and more general representation for the cross terms by taking into account the asynchronous signal propagation in distributed MIMO radar.

Another objective of this paper is to consider the impact of synchronization errors, or \emph{sync errors} for brevity, on target detection in distributed MIMO radar. Synchronization among TXs and RXs in distributed MIMO radar is non-trivial, as the sensors are spatially separated, driven by individual local clocks and oscillators. Phase synchronization, which is essential in applications requiring coherent processing such as direction finding, were considered in several studies. Specifically, the phase identifiability problem in self-calibrating MIMO radar was discussed in \cite{SunTangJunTSP14}. Various phase synchronization schemes involving centralized or distributed processing were proposed in \cite{YangBlum11}. A number of works examined signal detection \cite{AkcakayaNehoraiPhase10}, direction finding \cite{HeBlumCRB10}, imaging \cite{DingChen2015}, and beamforming \cite{ZhangVorobyov2016} in the presence of phase errors when timing/frequency errors are negligible. Meanwhile, \cite{HuLiuLiu2016} proposed a solution for target detection when the observations from different RXs are not correctly registered, i.e., aligned in the same spatial coordinate system, due to possible timing mismatches among the sensors. While these studies underscore the importance of synchronization, joint investigations of the effects of timing, frequency, and phase errors, which are coupled with each other, on distributed MIMO radar are lacking. Another limitation is that orthogonal waveforms are assumed in most cases.

\subsection{Main Contributions}
A systematic framework is presented in this paper, which covers signal
modeling, detection methods, and statistical analysis, for target
detection in distributed MIMO radar with non-orthogonal waveforms and
sync errors. The main contributions are summarized as follows.

\subsubsection{Modeling}
We develop a general asynchronous signal model for distributed MIMO
radar, which can incorporate timing, frequency, and phase errors among
RXs and TXs. The auto and cross terms in the MF output are the auto-
and cross-ambiguity functions of the waveforms sampled at distinct
delays and Doppler frequencies associated with individual propagation
paths. We show the model encompasses the co-located MIMO radar as a
special case, which will be used to benchmark the
distributed MIMO radar and shed light on the impact of cross terms and
sync errors on target detection.

\subsubsection{Detection Methods}
We consider coherent and non-coherent target detection methods for
distributed MIMO radar. We first briefly review a classical
non-coherent detector (NCD) \cite{HaimovichBlum08}, which is an energy
detector, and a recently introduced approximate coherent detector
(ACD) \cite{WangLi20RadarConf}, which phase-compensates the auto terms
in the MF output but neglects cross terms. We then introduce an
improved coherent detector (CD) that allows for cross terms and,
moreover, exploits diversities in signal strength among different
TX-RX paths. We also propose a new \emph{hybrid detector} (HD) as a
trade-off solution to bridge NCD and CD. HD coherently processes
output samples of each MF and non-coherently integrates across
different MFs. Since it requires phase coherence locally but not
across spatially distributed antennas, HD bypasses the stringent phase
synchronization requirement of CD and, meanwhile, enjoys additional
coherent processing gain over NCD.

\subsubsection{Analysis}
We provide a statistical analysis of the 4 detectors NCD, ACD, CD, and
HD. Closed-form expressions of the probability of false alarm and,
respectively, the probability of detection are derived for either
non-fluctuating or fluctuating targets. With simple tuning, these
expressions are applicable to cases with or without sync errors and,
furthermore, can be extended to co-located MIMO radar. Therefore, they
offer useful tools to investigate the impacts of asynchronous
propagation, cross terms, and sync errors in distributed MIMO
radar.

\subsubsection{Key Observations}
To examine the impact of non-orthogonal waveforms,
we consider different distributed set-ups with varying
offsets\footnote{Such offsets, caused by spatially distributed and
inherently asynchronous sensors, exist even if the RXs have perfect
timing, frequency and phase information of the TXs, and therefore
shall not be confused as sync errors.}
in delay, frequency, and phase among different TX-RX paths, which lead to
cross terms with varying magnitudes and phases compared with those of
the auto terms.
For benchmarking, we compare the distributed MIMO radar with a
co-located MIMO radar which employs orthogonal waveforms and contains
only auto terms in its MF outputs. This is possible since co-located
MIMO radar has co-located sensors which can operate synchronously
without delay, frequency, and phase offsets, thus obviating cross
terms. Our comparative study reveals that, interestingly, \emph{the
  non-orthogonal distributed MIMO radar may under- or out-performs the
  orthogonal co-located MIMO radar in target detection}. The
performance depends on if the cross terms behave as
\emph{interference}, which occur when the auto and cross terms add
\emph{destructively}, or as \emph{resources}, which occurs when they
add \emph{constructively}. It should be noted that delay, frequency,
and phase offsets, which are coupled with each other (e.g., a delay
offset implies a phase offset, see Section \ref{sec:system_model}),
all affect how cross terms add with auto terms.

Numerical simulations are also provided to illustrate the impact of
sync errors on the 4 detectors NCD, ACD, CD, and HD. Our results
indicate that: (a) NCD and HD are immune from phase errors, which
affect only coherent detectors ACD and CD; (b) frequency errors,
caused by either carrier offsets or Doppler estimation errors, affect
all but NCD; and (c) all detectors are affected by timing errors,
which cause the MF output to be sampled off the peak location of the
auto ambiguity function (see Section \ref{sec:system_model}), thus
resulting in a loss in the signal-to-noise ratio (SNR). If the timing
error is sufficiently small (i.e., much smaller than the reciprocal of
the waveform bandwidth but still significant relative to the carrier
period) so that the SNR loss is negligible, then it will only impact
coherent detectors such as ACD or CD as the timing-error-induced phase
error may not be negligible.

The remainder is organized as follows. A general signal model for
distributed MIMO radar is presented in Section
\ref{sec:system_model}. Target detection methods are discussed in
Section \ref{sec:detectors}, and their statistical analysis in Section
\ref{sec:performanceanalysis}. Section \ref{sec:simulation} contains
numerical results, followed by conclusions in Section
\ref{sec:conclusion}.

\emph{Notations}: We use boldface symbols for vectors (lower case) and
matrices (upper case). $(\cdot)^T$ denotes the transpose and
$(\cdot)^H$ the conjugate transpose. $\Vert\cdot\Vert$ and
$\vert\cdot\vert$ denote the vector 2-norm and absolute value,
respectively. $\mathbb{E}\{\cdot\}$ represents the statistical
expectation. $\mathcal{CN}(\ubf,\Sigmabf)$ denotes the complex
Gaussian distribution with mean $\ubf$ and covariance matrix
$\Sigmabf$. $[\Xbf]_{mm}$ indicates the $(m,m)$-th element of the
diagonal matrix $\Xbf$ while $[\xbf]_m$ denotes the $m$-th element of
the vector $\xbf$.

\begin{figure}[t]
\centering
\includegraphics[width=3in]{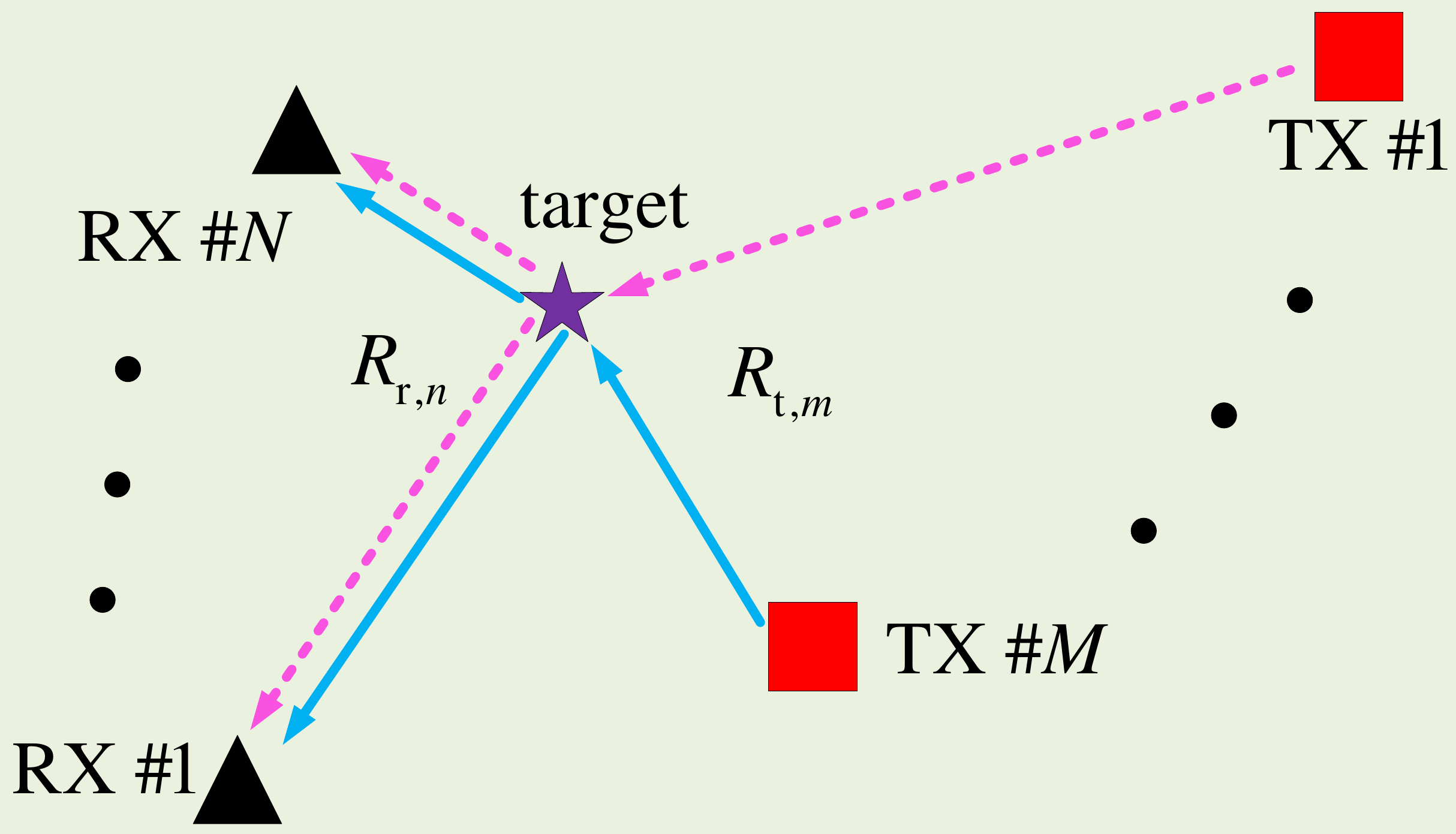}
\caption{Transmit and receive configuration of a distributed MIMO radar.}
\label{fig:configuration}
\end{figure}

\section{Signal Model}
\label{sec:system_model}

We first present a signal model for distributed MIMO radar with or without sync errors, and then briefly discuss the co-located MIMO radar, which is a special case of the former and will be employed as a benchmark for comparison.

\subsection{Distributed MIMO Radar}
\label{sec:system_model_distribute}

We consider a distributed MIMO radar system with $M$ TXs and $N$ RXs as shown in Fig.\,\ref{fig:configuration}. The TXs employ pulsed transmission to probe an area of interest by using $M$ waveforms. During a coherent processing interval, a succession of $K$
periodic pulses are transmitted by each TX. Specifically, at the $m$-th TX, the transmitted pulses are given by
\ben
\tilde{s}_m(t)=b_m u_m(t) e^{\jmath [2\pi (\hat{f}_\text{c}+\Delta_{\text{t},m}^{\text{c}}) t+\phi_{\text{t},m}]},
\een
where $u_m(t)=\sum_{k=0}^{K-1}p_{m}(t-k T_\text{s})$ is the baseband transmitted signal, $p_{m}(t)$ is the complex
envelope of a single pulse for TX $m$, $T_\text{s}$ is the pulse repetition
interval (PRI), $b_m$ is the transmit amplitude, $\hat{f}_\text{c}$ is
the nominal carrier frequency, $\Delta_{\text{t},m}^{\text{c}}$
denotes the carrier frequency error introduced by the $m$-th TX, and $\phi_{\text{t},m}$ is the carrier initial phase. The pulse waveform
$p_{m}(t)$ has unit energy and is of the same duration $T_\text{p}$ for all TXs. Therefore, $|b_{m}|^{2}$ denotes the energy transmitted in
a single pulse.

Suppose there is a moving target at a distance $R_{\text{t},m}$ to the $m$-th TX and a distance $R_{\text{r},n}$ to the $n$-th RX. The signal $\tilde s_n(t)$ observed at the $n$-th RX consists of
echoes from the target illuminated by $M$ waveforms
\begin{align}
\tilde s_n(t)
= &\sum\limits_{m=1}^{M}\alpha b_{m}\xi_{mn} u_m(t-\tau_{mn})
 \notag\\
 &\times e^{\jmath 2\pi (\hat{f}_\text{c}+\Delta_{\text{t},m}^{\text{c}}+\widetilde{f}_{mn}) (t-\tau_{mn})} e^{\jmath \phi_{\text{t},m}},
\end{align}
where $\alpha$ is the target amplitude, $\tau_{mn}=(R_{\text{t},m}+R_{\text{r},n})/c$ is the $(m,n)$-th
TX-RX propagation delay, and $\widetilde{f}_{mn}$ is the bistatic \emph{target Doppler frequency} \cite{HeLehmannBlum10,WangLiHimedNon11} observed by the $n$-th RX in response to the radar waveform transmitted from the $m$-th TX. In addition, $\xi_{mn}$ is the channel coefficient associated with the $(m,n)$-th TX-RX pair \cite{Richards05}:
\ben
\xi_{mn}=\sqrt{\frac{G_{\text{r},n}G_{\text{t},m}\lambda^2}{(4\pi)^3R_{\text{t},m}^2R_{\text{r},n}^2}},
\een
where $\lambda$ is the wavelength of the signal and $G_{\text{t},m}$ and $G_{\text{r},n}$ are the $m$-th TX and, respectively, $n$-th RX antenna gain.

A local carrier $e^{\jmath[2\pi(\hat{f}_\text{c}+\Delta_{\text{r},n}^{\text{c}})t+\phi_{\text{r},n}]}$ is generated at the $n$-th RX for down conversion, where $\Delta_{\text{r},n}^{\text{c}}$ and $\phi_{\text{r},n}$ denote the local carrier frequency error and initial phase, respectively. After down conversion, the baseband signal is
\begin{align}
&s_n(t) = \sum\limits_{m=1}^{M}\alpha
b_{m} \xi_{mn} u_m(t-\tau_{mn})e^{\jmath 2\pi(\hat{f}_\text{c}+\Delta_{\text{t},m}^\text{c}+\widetilde{f}_{mn})(t-\tau_{mn})}\notag\\&\times e^{\jmath \phi_{\text{t},n}}e^{-\jmath[2\pi(\hat{f}_\text{c}+\Delta_{\text{r},n}^{\text{c}})t+\phi_{\text{r},n}]}\notag\\
  &=\sum\limits_{m=1}^{M}\alpha
b_{m} \xi_{mn} u_m(t-\tau_{mn})e^{\jmath\psi_{mn}} \notag\\&\times e^{-\jmath 2\pi (\hat{f}_\text{c}+\Delta_{\text{r},n}^\text{c}) \tau_{mn}}
  e^{\jmath 2\pi f_{mn}(t-\tau_{mn})},
\end{align}
where $\psi_{mn}\triangleq\phi_{\text{t},m}-\phi_{\text{r},n}$ denotes the initial phase offset and $f_{mn}\triangleq\widetilde{f}_{mn}+\Delta_{\text{t},m}^{\text{c}}-\Delta_{\text{r},n}^{\text{c}}$ denotes the combined frequency offset between the $m$-th TX and $n$-th RX. A set of $M$ matched filters (MFs), each matched
to one of $M$ waveforms, are used at the $n$-th RX. Each MF requires estimates of the target delay $\tau_{mn}$ and Doppler $f_{mn}$ for compensation. In the following, we first consider the general case with possible sync errors, and then extend the result to the ideal case of no sync error, which is included as a benchmark for comparative studies.
\subsubsection{With Sync Errors}
At the $n$-th RX, $s_n(t)$ is convolved with $M$ MFs,
$g_{m}(t)=p_{m}^{*}(-t)e^{\jmath 2\pi
  (f_{mn}+\Delta_{mn}^{\text{f}})t}$, $m=1,\dots, M$, where
$\Delta_{mn}^{\text{f}}$ denotes the \emph{frequency
  error} between the
\emph{effective Doppler frequency} $f_{mn}$ and its estimate
$\hat{f}_{mn}$.\footnote{Although the frequency error
  $\Delta^{\text{f}}_{mn}$ includes both the carrier frequency error
  and Doppler mismatch, the former  is usually much smaller as the
  TXs and RXs are cooperative, which enables accurate tracking of the
  carrier frequency, e.g., via phase-locked loop. The Doppler mismatch
  can be more significant due to target motion uncertainty. Therefore,
  we may refer to $\Delta^{\text{f}}_{mn}$ as the frequency error or
  the Doppler error interchangeably.} Let
us define the \emph{cross ambiguity function} (CAF) as
\begin{align}\label{equ:AF}
\chi_{m\bar{m}}(\nu, f)=\int p_{m}(\mu) p_{\bar{m}}^*(\mu-\nu)e^{\jmath 2\pi f \mu}d\mu.
\end{align}
Then, the output of the $m$-th MF at the $n$-th RX $x_{mn}(t)$ can be written as
\begin{align}\label{eq:dataeq6}
&x_{mn}(t)
= \sum\limits_{\bar{m}=1}^{M}\alpha b_{\bar{m}} \xi_{\bar{m}n} e^{\jmath \psi_{\bar{m}n}}
e^{-\jmath 2\pi (\hat{f}_\text{c}+\Delta_{\text{r},n}^{\text{c}})\tau_{\bar{m}n}} e^{-\jmath 2\pi f_{\bar{m}n} \tau_{\bar{m}n}}\notag\\
&\times e^{\jmath2\pi (f_{mn}+\Delta_{mn}^{\text{f}})t}\sum_{k=0}^{K-1}\int
    p_{\bar{m}}(\mu-kT_\text{s}-\tau_{\bar{m}n})\notag\\
    &\times p_{m}^{*}(\mu-t)e^{\jmath 2\pi (f_{\bar{m}n}-f_{mn}-\Delta_{mn}^{\text{f}}) \mu}d\mu\notag\\
&= \sum\limits_{\bar{m}=1}^{M}\alpha b_{\bar{m}} \xi_{\bar{m}n}
e^{-\jmath 2\pi (\hat{f}_\text{c}+\Delta_{\text{r},n}^{\text{c}}) \tau_{\bar{m}n}}e^{\jmath2\pi (f_{mn}+\Delta_{mn}^{\text{f}})(t-\tau_{\bar{m}n})}\notag\\
&\times e^{\jmath \psi_{\bar{m}n}}\sum\limits_{k=0}^{K-1} \chi_{m\bar{m}}(t-\tau_{\bar{m}n}-k T_\text{s}, f_{\bar{m}n}-f_{mn}-\Delta_{mn}^{\text{f}})\notag\\
&\times
    e^{\jmath 2\pi k T_\text{s}( f_{\bar{m}n}-f_{mn}-\Delta_{mn}^{\text{f}})}.
\end{align}

The continuous-time signal $x_{mn}(t)$ is sampled at the pulse rate, leading to $K$ \emph{slow-time samples} obtained at time instants
$t=\tau_{mn}+\Delta_{mn}^{\text{t}}+k T_\text{s},\ k=0,\cdots, K-1$, where
$\Delta_{mn}^{\text{t}}$ denotes the \emph{timing error} between the true propagation delay $\tau_{mn}$ and its estimate $\hat{\tau}_{mn}$. Then,
the output samples can be written as
\begin{align}
&x_{mn}(k)
=x_{mn}(t)\Big|_{t=\tau_{mn}+\Delta_{mn}^{\text{t}}+kT_\text{s}} = \alpha b_{m}\xi_{mn}e^{\jmath 2\pi kT_\text{s}f_{mn}}
\notag\\&\times\chi_{mm}(\Delta_{mn}^{\text{t}}, -\Delta_{mn}^{\text{f}})e^{-\jmath 2\pi (\hat{f}_\text{c}+\Delta_{\text{r},n}^{\text{c}}) \tau_{mn}} e^{\jmath 2\pi(f_{mn}+\Delta_{mn}^{\text{f}})\Delta_{mn}^{\text{t}}}\notag\\
&\times e^{\jmath \psi_{mn}}+\sum\limits_{\bar{m}\not =m}\alpha b_{\bar{m}}\xi_{\bar{m}n} e^{\jmath \psi_{\bar{m}n}}e^{-\jmath 2\pi (\hat{f}_\text{c}+\Delta_{\text{r},n}^{\text{c}}) \tau_{\bar{m}n}}
e^{\jmath 2\pi k T_\text{s}f_{\bar{m}n}}\notag\\
&\times \chi_{m\bar{m}}(\tau_{mn}+\Delta_{mn}^{\text{t}}-\tau_{\bar{m}n}, f_{\bar{m}n}-f_{mn}-\Delta_{mn}^{\text{f}})\notag\\&\times e^{\jmath 2\pi(f_{mn}+\Delta_{mn}^{\text{f}})(\tau_{mn}+\Delta_{mn}^{\text{t}}-\tau_{\bar{m}n})},
\label{data cube}
\\
&m=1,\dots,M;\,n=1,\dots,N;\,k=0,\dots, K-1\notag.
\end{align}

\emph{Remark 1:} It can be seen that the output sample $x_{mn}(k)$
consists of $M$ components: the first term is the \emph{auto term}
between the $m$-th waveform and the $m$-th MF, and the other
components represent the \emph{cross terms} between the other $M-1$
waveforms and the $m$-th MF. The cross terms vanish when waveforms
$p_m(t)$ are orthogonal to each other, which is a routine assumption
in the MIMO literature. In practice, maintaining strict orthogonality
across time and frequency in distributed MIMO radar with asynchronous
propagation is infeasible \cite{AbramovichFrazer08}. With
non-orthogonal waveforms or waveforms that are orthogonal only with
zero delay/Doppler, cross terms are present as residuals, which may
become non-negligible and need to be accounted for.

\emph{Remark 2:} The derivation of
    \eqref{eq:dataeq6} and \eqref{data cube} appears to suggest that
    the radar receiver requires prior estimates of the target delay
    and Doppler, which are unnecessary. The problem is addressed by
    having the receiver scanning through the delay/Doppler uncertainty
    region, which is discretized into a set of range/Doppler bins. In
    our derivation, range/Doppler measurements are obtained by using a
    set of MFs, each matched to a distinct Doppler frequency, and
    sampling the MF outputs at the Nyquist rate. In practice, the
    above process is often approximated by a more efficient procedure,
    which involves processing the target return using a fixed MF,
    sampling the MF output in fast- and slow-times, and then
    converting to the frequency domain by the fast Fourier transform
    (FFT) \cite{Richards05}. Note that \eqref{data cube} describes the
    observed signal only for the range-Doppler bin with the
    target. For non-target range-Doppler bins, the measurements
    contain noise. These two types of measurements are summarized by
    the hypothesis testing data model in \eqref{equ:hypothese}.

Next, we stack the $K$ slow-time samples and form $\xbf_{mn}=[x_{mn}(0),\cdots,x_{mn}(K-1)]^T$, which can be expressed as
\begin{align}\label{equ:abservationerror}
\xbf_{mn}=\alpha\Sbf_n\Xcalbf_{mn}\hbf_{mn},
\end{align}
where the $K\times M$ Doppler steering matrix $\Sbf_n$ is
\begin{align}\label{eq:Sn}
\Sbf_n&=[\sbf(f_{1n}),\cdots,\sbf(f_{Mn})],\\
\sbf(f)&=[1,e^{\jmath 2\pi T_\text{s}f},\cdots,e^{\jmath 2\pi (K-1)T_\text{s}f}]^T,\notag
\end{align}
the $M\times M$ ambiguity function matrix $\Xcalbf_{mn}$ is diagonal with diagonal elements given by
\begin{align}\label{eq:errorX}
[\Xcalbf_{mn}]_{\bar{m}\bar{m}}=\chi_{m\bar{m}}(\tau_{mn}+\Delta_{mn}^{\text{t}}-\tau_{\bar{m}n},f_{\bar{m}n}-f_{mn}-\Delta_{mn}^{\text{f}}),
\end{align}
and the $\bar{m}$-th element of the $M\times 1$ channel vector $\hbf_{mn}$ is
\begin{align}\label{eq:errorh}
[\hbf_{mn}]_{\bar{m}}&=b_{\bar{m}}\xi_{\bar{m}n} e^{\jmath\psi_{\bar{m}n}}e^{-\jmath2\pi (\hat{f}_\text{c}+\Delta_{\text{r},n}^{\text{c}})\tau_{\bar{m}n}}\notag\\&\times e^{\jmath 2\pi(f_{mn}+\Delta_{mn}^{\text{f}})(\tau_{mn}+\Delta_{mn}^{\text{t}}-\tau_{\bar{m}n})}.
\end{align}

\subsubsection{Without Sync Errors}
\Cref{data cube,equ:abservationerror,eq:Sn,eq:errorX,eq:errorh}, which
describe the general signal model for distributed MIMO radar, also
hold for the ideal case without sync errors, by setting
$\Delta_{mn}^{\text{f}}=0$, $\Delta_{\text{r},n}^{\text{c}}=0$, and $\Delta_{mn}^{\text{t}}=0$, $\forall m,\ \forall n$. In other words, we need replace \eqref{eq:errorX} and \eqref{eq:errorh} by
\begin{gather}\label{equ:noerrorX}
[\Xcalbf_{mn}]_{\bar{m}\bar{m}}=\chi_{m\bar{m}}(\tau_{mn}-\tau_{\bar{m}n},f_{\bar{m}n}-f_{mn}),
\\
\label{equ:noerrorh}
[\hbf_{mn}]_{\bar{m}}=b_{\bar{m}}\xi_{\bar{m}n} e^{\jmath\psi_{\bar{m}n}}e^{-\jmath2\pi \hat{f}_\text{c}\tau_{\bar{m}n}}e^{\jmath 2\pi f_{mn}(\tau_{mn}-\tau_{\bar{m}n})}.
\end{gather}

\emph{Remark 3:} It is worth to note that matched filtering and sampling only require the knowledge of the delay $\tau_{mn}$ and Doppler $f_{mn}$, but not the carrier phase offset $\psi_{mn}$. Therefore, phase errors are absent from the signal model \Cref{data cube,equ:abservationerror,eq:Sn,eq:errorX,eq:errorh,equ:noerrorX,equ:noerrorh}. However, for coherent detection, the observed signal $\xbf_{mn}$ will be phase-compensated, and phase estimation errors will impact such detectors (see Sections \ref{subsec:CD} and \ref{subsec:performanceCD} for details).
\subsection{Co-Located MIMO Radar}
\label{subsec:colocatedMIMO}
Co-located MIMO radar, which is a special case of distributed MIMO radar, can be described by \Cref{data cube,equ:abservationerror,eq:Sn,eq:errorX,eq:errorh,equ:noerrorX,equ:noerrorh} with some simplifications. Specifically, with co-located antennas, we have identical target delay $\tau_{mn}$, $\forall m,n$, and identical Doppler frequency $f_{mn}$, $\forall m,n$. In addition, if the TXs share the same oscillator, and so are the RXs, then the phase offset $\psi_{mn}$ is constant $\forall m,n$.

Assume the radar employs waveforms that are orthogonal with zero delay and Doppler, i.e.,
\begin{align}
\chi_{m\bar{m}}(0,0)=0,\ \forall m\neq\bar{m}.
\end{align}
Then, the cross terms in \eqref{data cube} disappear for co-located MIMO radar when there is no sync error. In this case, \eqref{data cube} can be simplified as
\begin{align}
&x_{mn}(k)
=\alpha b_{m}\xi_{mn}
e^{-\jmath 2\pi \hat{f}_\text{c} \tau_{mn}}\chi_{mm}(0, 0) e^{\jmath 2\pi kT_\text{s}f_{mn}}
e^{\jmath \psi_{mn}}.
\label{data cube_colocated}
\end{align}
In turn, \eqref{eq:errorX} and \eqref{eq:errorh} reduces to
\begin{gather}
\label{eq:comimoX}
[\Xcalbf_{mn}]_{\bar{m}\bar{m}}=
\begin{cases}
   0, & \bar{m}\neq m, \\
\\
\chi_{mm}(0,0), &\bar{m}= m,
\end{cases}
\\
\label{eq:comimoh}
[\hbf_{mn}]_{\bar{m}}=b_{\bar{m}}\xi_{\bar{m}n}
e^{-\jmath 2\pi \hat{f}_\text{c} \tau_{\bar{m}n}}
e^{\jmath \psi_{\bar{m}n}}.
\end{gather}
\Cref{eq:comimoX,eq:comimoh} along with \eqref{eq:Sn}, describe the \emph{orthogonal co-located MIMO radar}, which will be employed to benchmark the non-orthogonal distributed MIMO radar.
\section{Target Detection}
\label{sec:detectors}
Let $\ybf_{mn}$ denote the noise contaminated observation of $\xbf_{mn}$. The target detection problem is described by the following hypothesis testing:
\begin{align}\label{equ:hypothese}
\mathcal{H}_0: \ybf_{mn}&=\wbf_{mn},\notag\\
\mathcal{H}_1: \ybf_{mn}&=\alpha\Sbf_n\Xcalbf_{mn}\hbf_{mn}+\wbf_{mn},\\
&m=1,2,\cdots,M,\ n=1,2,\cdots,N,\notag
\end{align}
where $\wbf_{mn}$ is the noise, assumed to be Gaussian distributed, $\wbf_{mn}\sim\mathcal{CN}(\mathbf{0},\sigma_{mn}^2\Ibf)$. Note that the above hypothesis testing applies to both distributed and co-located MIMO radars. In the following, we first consider target detection approaches for the general case, i.e., distributed MIMO radar with possible sync errors, and then extend/simplify the solutions to the cases with no sync error and co-located MIMO radar. For target detection, we discuss several detectors, including a conventional non-coherent detector (NCD) \cite{HaimovichBlum08}, an approximate coherent detector (ACD) \cite{WangLi20RadarConf}, a coherent detector (CD), and a hybrid detector (HD). The latter two are new.
\subsection{Non-Coherent Detector}
\label{sub:NCD}
A simple detector for the hypothesis testing \eqref{equ:hypothese} is based on non-coherent integration of the MF outputs \cite{HaimovichBlum08}:
\begin{align}\label{eq:NCD}
T_\text{NCD}\triangleq \sum_{m=1}^M\sum_{n=1}^N\ybf_{mn}^H\ybf_{mn}\gs_ {\mathcal{H}_0}^{\mathcal{H}_1}\gamma_\text{NCD},
\end{align}
where $\gamma_\text{NCD}$ is a threshold set for a given level of false alarm. It is clear that the above NCD is an energy detector.

\subsection{Coherent Detectors}
\label{subsec:CD}
The above NCD does not require any phase synchronization. Improved detection performance can be achieved by exploiting phase information. One such detector, ACD, was introduced in \cite{WangLi20RadarConf}, which performs phase compensation for the auto terms in the MF output \eqref{data cube}. Specifically, let $\hat{\psi}_{mn}$, $\hat{\tau}_{mn}$, and $\hat{f}_{mn}$ denote estimates of the phase offset, delay, and Doppler frequency. The ACD is given by
\begin{align}\label{eq:teststatisticACD}
T_{\text{ACD}}=\left\vert\sum_{m=1}^M\sum_{n=1}^N\sum_{k=0}^{K-1} e^{-\jmath\hat{\theta}_{mnk}}y_{mn}(k)\right\vert^2\gs_ {\mathcal{H}_0}^{\mathcal{H}_1}\gamma_{\text{ACD}},
\end{align}
where $\gamma_{\text{ACD}}$ is the threshold, $y_{mn}(k)$ denotes the $k$-th element of $\ybf_{mn}$ and
\begin{gather}\label{eq:thetaest}
\hat{\theta}_{mnk}=\hat{\psi}_{mn}-2\pi \hat{f}_\text{c}\hat{\tau}_{mn}+2\pi kT_\text{s}\hat{f}_{mn},
\\
\label{eq:phaseest}
\hat{\psi}_{mn}\triangleq\psi_{mn}+\Delta_{mn}^{\text{p}},
\\
\label{eq:delayest}
\hat{\tau}_{mn}\triangleq\tau_{mn}+\Delta_{mn}^{\text{t}},
\\
\label{eq:Dopplerest}
\hat{f}_{mn}\triangleq f_{mn}+\Delta_{mn}^{\text{f}},
\end{gather}
where $\Delta_{mn}^{\text{p}}$, $\Delta_{mn}^{\text{t}}$, and $\Delta_{mn}^{\text{f}}$ denote the phase, timing, and Doppler errors.

Albeit simple, the ACD has two limitations. First, it performs phase compensation only for the auto-term, while neglecting the cross terms in \eqref{data cube}, which is non-negligible when the waveforms are not orthogonal. Second, it applies equal weights in combining the outputs from different MFs, which is suboptimal since the TX-RX propagation paths associated with different MFs are different with potentially different SNRs. To address these limitations, we propose an improved CD with derivation presented in Appendix \ref{appendexCD}. Specifically, let $\hat{\Sbf}_n$, $\hat{\Xcalbf}_{mn}$, and $\hat{\hbf}_{mn}$ be formed as in \Cref{eq:Sn,eq:errorX,eq:errorh}, by using the phase, delay, and Doppler frequency estimates:
\begin{gather}\label{eq:estimatedDoppler}
\hat{\Sbf}_n=[\sbf(\hat{f}_{1n}),\cdots,\sbf(\hat{f}_{Mn})],
\\
\label{equ:timingerrorAF}
[\hat{\Xcalbf}_{mn}]_{\bar{m}\bar{m}}=\chi_{m\bar{m}}(\hat{\tau}_{mn}-\hat{\tau}_{\bar{m}n},\hat{f}_{\bar{m}n}-\hat{f}_{mn}),
\\
\label{eq:estimatedphase}
[\hat{\hbf}_{mn}]_{\bar{m}}=b_{\bar{m}}\xi_{\bar{m}n} e^{\jmath\hat{\psi}_{\bar{m}n}}e^{-\jmath2\pi \hat{f}_\text{c}\hat{\tau}_{\bar{m}n}}e^{\jmath 2\pi\hat{f}_{mn}(\hat{\tau}_{mn}-\hat{\tau}_{\bar{m}n})}.
\end{gather}
Then, the new CD is given by
\begin{align}\label{eq:CD}
T_\text{CD}=\left\vert\sum_{m=1}^M\sum_{n=1}^N(\hat{\Sbf}_n\hat{\Xcalbf}_{mn}\hat{\hbf}_{mn})^H\ybf_{mn}\right\vert^2\gs_ {\mathcal{H}_0}^{\mathcal{H}_1}\gamma_\text{CD},
\end{align}
where $\gamma_\text{CD}$ denotes the test threshold. It can be seen that the CD sequentially performs Doppler filtering by $\hat{\Sbf}_n$, joint phase compensation and amplitude weighting by $\hat{\Xcalbf}_{mn}$ and $\hat{\hbf}_{mn}$, followed by coherent integration across antennas.

\subsection{Hybrid Detector}
The above CD requires the knowledge of the phases, the CAFs of all waveforms, and the channel coefficients $\xibf_{mn}$. Although achieving the best performance with ideal knowledge, CD is sensitive to knowledge/estimation errors. In Appendix \ref{appendexHD}, we derive a new HD detector that bypasses the stringent requirement of CD and can still achieve considerable improvement over the NCD, i.e., it offers a compromise between CD and NCD. Specifically, the HD is given by
\begin{align}\label{eq:HD}
T_\text{HD}=\sum_{m=1}^M\sum_{n=1}^N\Vert\hat{\Sbf}_n(\hat{\Sbf}_n^H\hat{\Sbf}_n)^{-1}\hat{\Sbf}_n^H\ybf_{mn}\Vert^2\gs_ {\mathcal{H}_0}^{\mathcal{H}_1}\gamma_\text{HD},
\end{align}
where it is assumed $K\geq M$ so that the matrix inverse exists.
Clearly, HD projects $\ybf_{mn}$ onto the subspace spanned by the Doppler steering vectors $\hat{\Sbf}_n$, which is coherent processing of the signal observed at the $(m,n)$-th MF, followed by non-coherent integration across different RXs and TXs. Hence, it is a $\emph{hybrid}$ detector. The projection preserves the target signal $\ybf_{mn}$ undistorted, while rejecting the noise component in the orthogonal subspace. This leads to an improved SNR, allowing HD to outperform the NCD. Note that HD requires phase coherent only locally, i.e., within the output of each MF, but not across spatially distributed antennas.

\subsection{Impact of Sync Errors}
\label{subsec:errors}
For NCD, sync errors in timing, Doppler, and phase only affect the MF measurements in \eqref{equ:hypothese}, but do not have any impact on the NCD implementation \eqref{eq:NCD}. This implies that without sync errors, NCD is still given by \eqref{eq:NCD}, except that the MF outputs $\ybf_{mn}$ are represented by \eqref{equ:hypothese}, along with \eqref{eq:Sn}, \eqref{equ:noerrorX}, and \eqref{equ:noerrorh}.

For ACD and CD, sync errors affect both measurements and implementation, due to the additional phase/amplitude compensation employed by the latter. In the case of no sync errors, the MF measurements are given by \eqref{equ:hypothese}, along with \eqref{eq:Sn}, \eqref{equ:noerrorX}, and \eqref{equ:noerrorh}. The phase compensations $\{\hat{\theta}_{mnk}\}$ in \eqref{eq:thetaest} should be replaced by their true values for ACD \eqref{eq:teststatisticACD}, while for CD \eqref{eq:CD}, the amplitude/phase compensation quantities $\{\hat{\Sbf}_{n}\}$, $\{\hat{\Xcalbf}_{mn}\}$, and $\{\hat{\hbf}_{mn}\}$ should be replaced by their error-free counterparts.

For HD, sync errors also affect both the measurements and implementation due to the Doppler projection in \eqref{eq:HD}. In the absence of sync errors, the measurements \eqref{equ:hypothese} have the same representation as discussed above, whereas \eqref{eq:HD} should be implemented with the true Doppler matrices $\{\Sbf_n\}$.
\subsection{Extension to Co-Located MIMO Radar}
\label{subsec:coMIMOdetection}
For co-located MIMO radar with orthogonal waveforms, the MF measurements are represented by \eqref{equ:hypothese}, along with \eqref{eq:Sn}, \eqref{eq:comimoX}, and \eqref{eq:comimoh}. The implementations of the NCD, ACD/CD, and HD remain the same as in distributed MIMO radar, which are given by \eqref{eq:NCD}, \eqref{eq:teststatisticACD}, \eqref{eq:CD} and \eqref{eq:HD}, respectively. Finally, the discussions on the impact of sync errors in Section \ref{subsec:errors} for distributed MIMO radar are also applicable to co-located MIMO radar.
\section{Statistical Analysis}
\label{sec:performanceanalysis}
In this section, we provide a statistical analysis of the 4 detectors, NCD, CD, ACD, and HD, introduced in Section \ref{sec:detectors} and derive expressions of their false alarm and detection probabilities. We only consider the general case with sync errors, since the test statistic of each detector retains the same form, although the statistical distributions are affected by the presence/absence of sync errors, but can be easily determined by using the corresponding signal representations as detailed in Section \ref{subsec:errors}. For each detector, we derive the probability of detection for the case of \emph{non-fluctuating} target and, respectively, the case of \emph{fluctuating} target assuming the Swerling \uppercase\expandafter{\romannumeral1} model \cite{Richards05}. Finally, we briefly discuss how to extend these results to co-located MIMO radar.
\subsection{NCD}
\begin{Theorem}\ThmLabel{theoremNCD}
Given the signal model \eqref{equ:hypothese} with or without sync errors, the probability of false alarm and the probability of detection with a non-fluctuating target for the NCD \eqref{eq:NCD} are given by
\begin{gather}\label{eq:falseNCD}
P_f=\frac{\Gamma(KMN)-\bar{\Gamma}(KMN,\gamma_\text{NCD}/\sigma^2)}{\Gamma(KMN)},
\\
\label{eq:detectionNCD}
P_d=Q_{KMN}\Big(\sqrt{\lambda_\text{NCD}},\sqrt{\frac{2\gamma_\text{NCD}}{\sigma^2}}\Big),
\end{gather}
where $\bar{\Gamma}(\cdot,\cdot)$ denotes the upper incomplete Gamma function, $\Gamma(\cdot)$ the Gamma function, $Q_m(a,b)$ the generalized Marcum-Q function, and the noncentrality parameter is given by
\ben\label{eq:lambdaNCD}
\lambda_\text{NCD}=\sum_{m=1}^M\sum_{n=1}^N\frac{2\vert\alpha\vert^2}{\sigma^2}\Vert\Sbf_n\Xcalbf_{mn}\hbf_{mn}\Vert^2.
\een
\end{Theorem}
\begin{IEEEproof}
See Appendix \ref{appendexAnalysisNCD}.
\end{IEEEproof}

Next, we examine the average probability of detection in the case of fluctuating target. For Swerling \uppercase\expandafter{\romannumeral1} target, the probability density function (pdf) of the radar cross section (RCS) $\rho=\vert\alpha\vert^2$ is \cite{Richards05}
\ben
f(\rho)=\frac{1}{\bar{\rho}}e^{-\rho/\bar{\rho}},
\een
where $\bar{\rho}=\mathbb{E}\{\rho\}$. Let $\lambda'_\text{NCD}\triangleq\lambda_\text{NCD}/\rho$. The average probability of detection is given by
\begin{align}\label{eq:averageNCD}
&\bar{P}_d=\int_{0}^{\infty}f(\rho)P_dd\rho\notag\\
&=\int_{0}^{\infty}\frac{1}{\bar{\rho}}e^{-\rho/\bar{\rho}}Q_{KMN}\Big(\sqrt{\lambda'_\text{NCD}\rho},\sqrt{\frac{2\gamma_\text{NCD}}{\sigma^2}}\Big)d\rho\notag\\
&=\frac{\bar{\Gamma}(KMN,\gamma_\text{NCD}/\sigma^2)}{\Gamma(KMN)}\notag\\&+\frac{\lambda'_\text{NCD}\big(\frac{\gamma_\text{NCD}}{\sigma^2}\big)^{KMN}F_1(1,KMN+1,\frac{\gamma_\text{NCD}\lambda'_\text{NCD}}{\sigma^2(\lambda'_\text{NCD}+2/\bar{\rho})})}{(KMN)!(\lambda'_\text{NCD}+2/\bar{\rho})e^{\frac{\gamma_\text{NCD}}{\sigma^2}}},
\end{align}
where $x!$ denotes the factorial, $F_1(\cdot,\cdot,\cdot)$ is the Kummer confluent hypergeometric function \cite{prudnikov1986integrals}, and the third equality is obtained by using \cite[Theorem 1]{Sofotasios2014}.
\subsection{CD}
\label{subsec:performanceCD}
\begin{Theorem}\ThmLabel{theoremDCD}
Given the signal model \eqref{equ:hypothese} with or without sync errors, the probability of false alarm and the probability of detection with a non-fluctuating target for the CD \eqref{eq:CD} are given by
\begin{gather}\label{eq:falseDCD}
P_f=e^{-\frac{\gamma_\text{CD}}{\varsigma\sigma^2}},
\\
\label{eq:detectionDCD}
P_d=Q_1(\sqrt{\lambda_\text{CD}},\sqrt{\frac{2\gamma_\text{CD}}{\varsigma\sigma^2}}),
\end{gather}
where the noncentrality parameter is given by
\ben\label{eq:lambdaDCD}
\lambda_\text{CD}=\frac{2\vert\alpha\vert^2\left\vert\sum_{m=1}^M\sum_{n=1}^N(\hat{\Sbf}_n\hat{\Xcalbf}_{mn}\hat{\hbf}_{mn})^H\Sbf_n\Xcalbf_{mn}\hbf_{mn}\right\vert^2}{\sigma^2\sum_{m=1}^M\sum_{n=1}^N\Vert\hat{\Sbf}_n\hat{\Xcalbf}_{mn}\hat{\hbf}_{mn}\Vert^2},
\een
and the scaling factor is
\ben\label{eq:varsigma}
\varsigma=\sum_{m=1}^M\sum_{n=1}^N\Vert\hat{\Sbf}_n\hat{\Xcalbf}_{mn}\hat{\hbf}_{mn}\Vert^2.
\een
\end{Theorem}
\begin{IEEEproof}
See Appendix \ref{appendexAnalysisDCD}.
\end{IEEEproof}

Similar to NCD, the average probability of detection for the case of Swerling \uppercase\expandafter{\romannumeral1} fluctuating target is given by
\begin{align}\label{eq:averageCD}
&\bar{P}_d=\int_{0}^{\infty}\frac{1}{\bar{\rho}}e^{-\rho/\bar{\rho}}Q_1(\sqrt{\lambda'_\text{CD}\rho},\sqrt{\frac{2\gamma_\text{CD}}{\varsigma\sigma^2}})d\rho\notag\\
&=\bar{\Gamma}(1,\frac{\gamma_\text{CD}}{\varsigma\sigma^2})+\frac{\lambda'_\text{CD}\gamma F_1(1,2,\frac{\gamma_\text{CD}\lambda'_\text{CD}}{\varsigma\sigma^2(\lambda'_\text{CD}+2/\bar{\rho})})}{\varsigma\sigma^2(\lambda'_\text{CD}+2/\bar{\rho})e^{\frac{\gamma_\text{CD}}{\varsigma\sigma^2}}},
\end{align}
where $\lambda'_\text{CD}\triangleq\lambda_\text{CD}/\rho$.
\subsection{ACD}
\begin{Theorem}\ThmLabel{theoremACD}
Given the signal model \eqref{equ:hypothese} with or without sync errors, the probability of false alarm and the probability of detection with a non-fluctuating target for the ACD \eqref{eq:teststatisticACD} are given by
\begin{gather}\label{eq:falseACD}
P_f=e^{-\frac{\gamma_\text{ACD}}{KMN\sigma^2}},
\\
\label{eq:detectionACD}
P_d=Q_1(\sqrt{\lambda_\text{ACD}},\sqrt{\frac{2\gamma_\text{ACD}}{KMN\sigma^2}}),
\end{gather}
where
\ben\label{eq:lambdaHCD}
\lambda_\text{ACD}=\frac{2\left\vert\sum_{m=1}^M\sum_{n=1}^N\sum_{k=0}^{K-1}e^{-\jmath \hat{\theta}_{mnk}}x_{mn}(k)\right\vert^2}{KMN\sigma^2},
\een
where $\hat{\theta}_{mnk}$ is defined in \eqref{eq:thetaest} and $x_{mn}(k)$ is defined in \eqref{data cube}.
\end{Theorem}
\begin{IEEEproof}
See Appendix \ref{appendexAnalysisACD}.
\end{IEEEproof}

Define $\lambda'_\text{ACD}\triangleq\lambda_\text{ACD}/\rho$. The average probability of detection for the case of Swerling \uppercase\expandafter{\romannumeral1} fluctuating target is given by
\begin{align}\label{eq:averageACD}
&\bar{P}_d=\bar{\Gamma}(1,\frac{\gamma_\text{ACD}}{KMN\sigma^2})\notag\\&+\frac{\lambda'_\text{ACD}\gamma_\text{ACD} F_1(1,2,\frac{\gamma_\text{ACD}\lambda'_\text{ACD}}{KMN\sigma^2(\lambda'_\text{ACD}+2/\bar{\rho})})}{KMN\sigma^2(\lambda'_\text{ACD}+2/\bar{\rho})e^{\frac{\gamma_\text{ACD}}{KMN\sigma^2}}}.
\end{align}
\subsection{HD}
\begin{Theorem}\ThmLabel{theoremHD}
Given the signal model \eqref{equ:hypothese} with or without sync errors, the probability of false alarm and the probability of detection with a non-fluctuating target for the HD \eqref{eq:HD} are given by
\begin{gather}\label{eq:falseHD}
P_f=\frac{\Gamma(NM^2)-\bar{\Gamma}(NM^2,\gamma_\text{HD}/\sigma^2)}{\Gamma(NM^2)},
\\
\label{eq:detectionHD}
P_d=Q_{NM^2}\Big(\sqrt{\lambda_\text{HD}},\sqrt{\frac{2\gamma_\text{HD}}{\sigma^2}}\Big),
\end{gather}
where the noncentrality parameter is given by
\begin{align}\label{eq:lambdaHD}
\lambda_\text{HD}=\frac{\vert\alpha\vert^2\sum_{m=1}^M\sum_{n=1}^N\Vert\hat{\Sbf}_n(\hat{\Sbf}_n^H\hat{\Sbf}_n)^{-1}\hat{\Sbf}_n^H\Sbf_n\Xcalbf_{mn}\hbf_{mn}\Vert^2}{\sigma^2/2}.
\end{align}
\end{Theorem}
\begin{IEEEproof}
See Appendix \ref{appendexAnalysisHD}.
\end{IEEEproof}

Then, the average probability of detection for the case of Swerling \uppercase\expandafter{\romannumeral1} fluctuating target is given by
\begin{align}\label{eq:averageHD}
\bar{P}_d=&\frac{\bar{\Gamma}(NM^2,\gamma_\text{HD}/\sigma^2)}{\Gamma(NM^2)}\notag\\&+\frac{\lambda'_\text{HD}\big(\frac{\gamma_\text{HD}}{\sigma^2}\big)^{NM^2}F_1(1,NM^2+1,\frac{\gamma_\text{HD}\lambda'_\text{HD}}{\sigma^2(\lambda'_\text{HD}+2/\bar{\rho})})}{(NM^2)!(\lambda'_\text{HD}+2/\bar{\rho})e^{\frac{\gamma_\text{HD}}{\sigma^2}}},
\end{align}
where $\lambda'_\text{HD}\triangleq\lambda_\text{HD}/\rho$.
\subsection{Extension to Co-Located MIMO Radar}
As discussed in Section \ref{subsec:coMIMOdetection}, NCD, CD, ACD, and HD can be applied for target detection in co-located MIMO radar. It is easy to see that Theorems \ThmRef{theoremNCD} to \ThmRef{theoremHD}, as well as the expressions \eqref{eq:averageNCD}, \eqref{eq:averageCD}, \eqref{eq:averageACD}, and \eqref{eq:averageHD} for the average probability of detection, still hold for co-located MIMO radar. The only difference is that the noncentrality parameter $\lambda$ involved in each detector, as well as $\varsigma$ \eqref{eq:varsigma} for the CD, should be calculated by using the corresponding signal representations, as discussed in Section \ref{subsec:coMIMOdetection}.
\section{Simulation Results}
\label{sec:simulation}
In this section, simulation results are presented to demonstrate the performance of the NCD \cite{HaimovichBlum08}, ACD \cite{WangLi20RadarConf}, along with the proposed CD and HD, for target detection in distributed MIMO radar. The performance of these detectors are assessed by using both computer simulation and the analytical results reported in Section \ref{sec:performanceanalysis}. The SNR of the $(m,n)$-th propagation path, which is measured at the $n$-th RX matched to the $m$-th TX waveform, is defined as
\begin{align}
\text{SNR}_{mn}=\frac{\vert b_m\xi_{mn}\vert^2\mathbb{E}\{\vert\alpha\vert^2\}}{\sigma_{mn}^2},
\end{align}
where the noise variance is chosen as $\sigma_{mn}^2=1$. We consider a Swerling \uppercase\expandafter{\romannumeral1} target mdel, where the target amplitude $\alpha\sim\mathcal{CN}(0,\sigma^2)$ is randomly generated from trial to trial but remains fixed within a coherent processing interval (CPI) in Monte Carlo simulations. We assume identical SNR for all paths, i.e., $\text{SNR}_{mn}=\text{SNR}$, $\forall m,n$, except in Section \ref{subsec:diverseresults} where the effect of different SNRs is examined. The simulation scenarios involve a distributed MIMO radar with $M=2$ TXs and $N=1$ RX. The propagation delays are $\tau_{11}=0.61T_\text{p}$ and $\tau_{21}=0.1T_\text{p}$ unless otherwise stated, where $T_\text{p}=10^{-5}$ s is the pulse duration. The pulse repetition frequency (PRF) is 500 Hz, the carrier frequency is 3 GHz, the target Doppler frequencies are $f_{11}=200$ Hz and $f_{21}=190$ Hz, unless otherwise stated, and the number of pulses within a CPI is $K=12$. The phases are $\psi_{11}=0.1\pi$ and $\psi_{21}=0.3\pi$ unless otherwise stated and the probability of false alarm is $P_f=10^{-4}$.

In the following, we first introduce two sets of linear frequency modulation (LFM) based waveforms, also known as chirps, which are used by the MIMO radar for testing. Then, we examine the performance of these detectors in various distributed environments with non-identical propagation path strengths, different delays, phases, and Doppler frequencies, as well as in the presence of sync errors.
\subsection{Test Waveforms}
\begin{figure}[t]
\centering
\bt{c}
\includegraphics[width=2.4in]{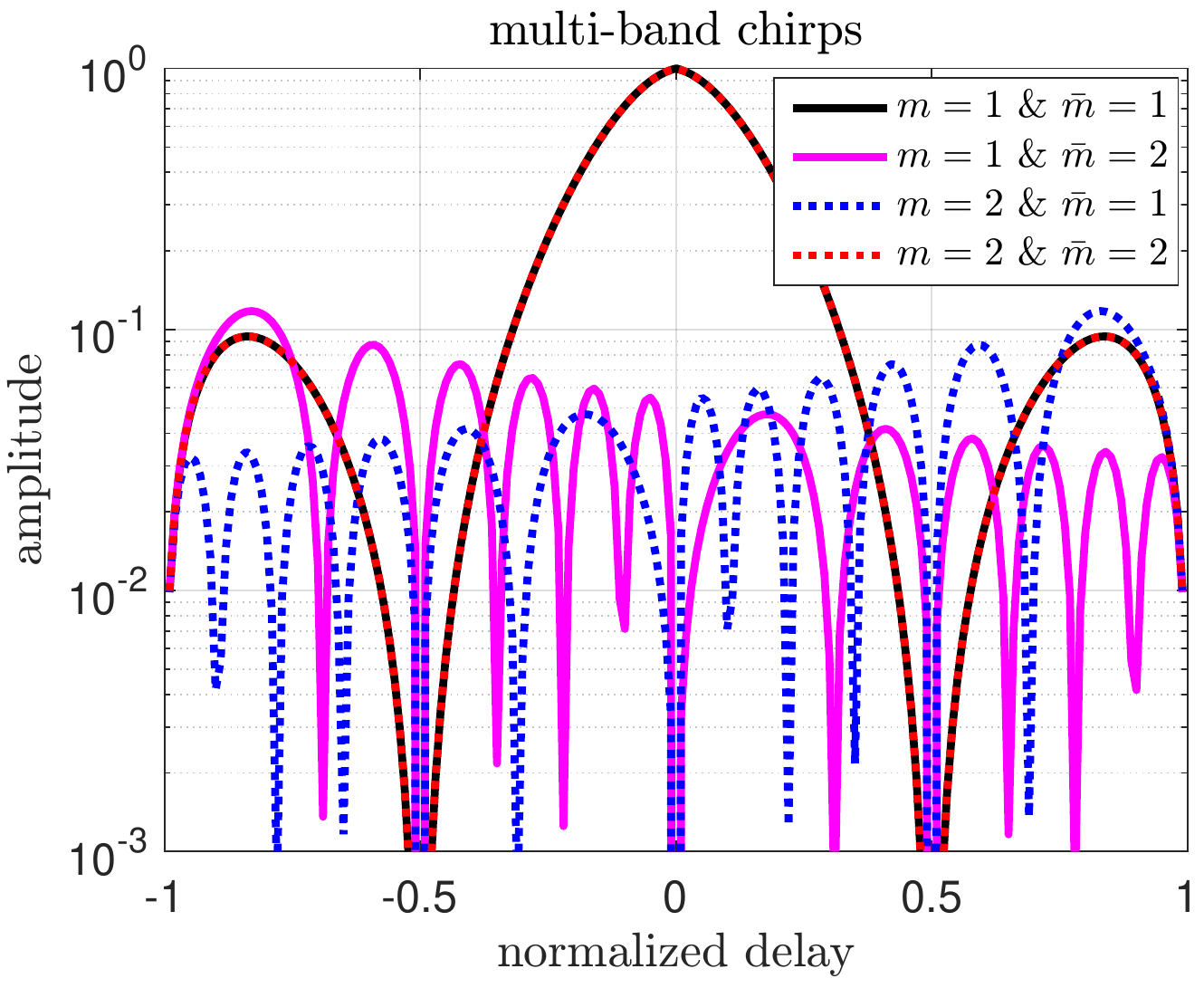}\\
(a)\\ 
\includegraphics[width=2.4in]{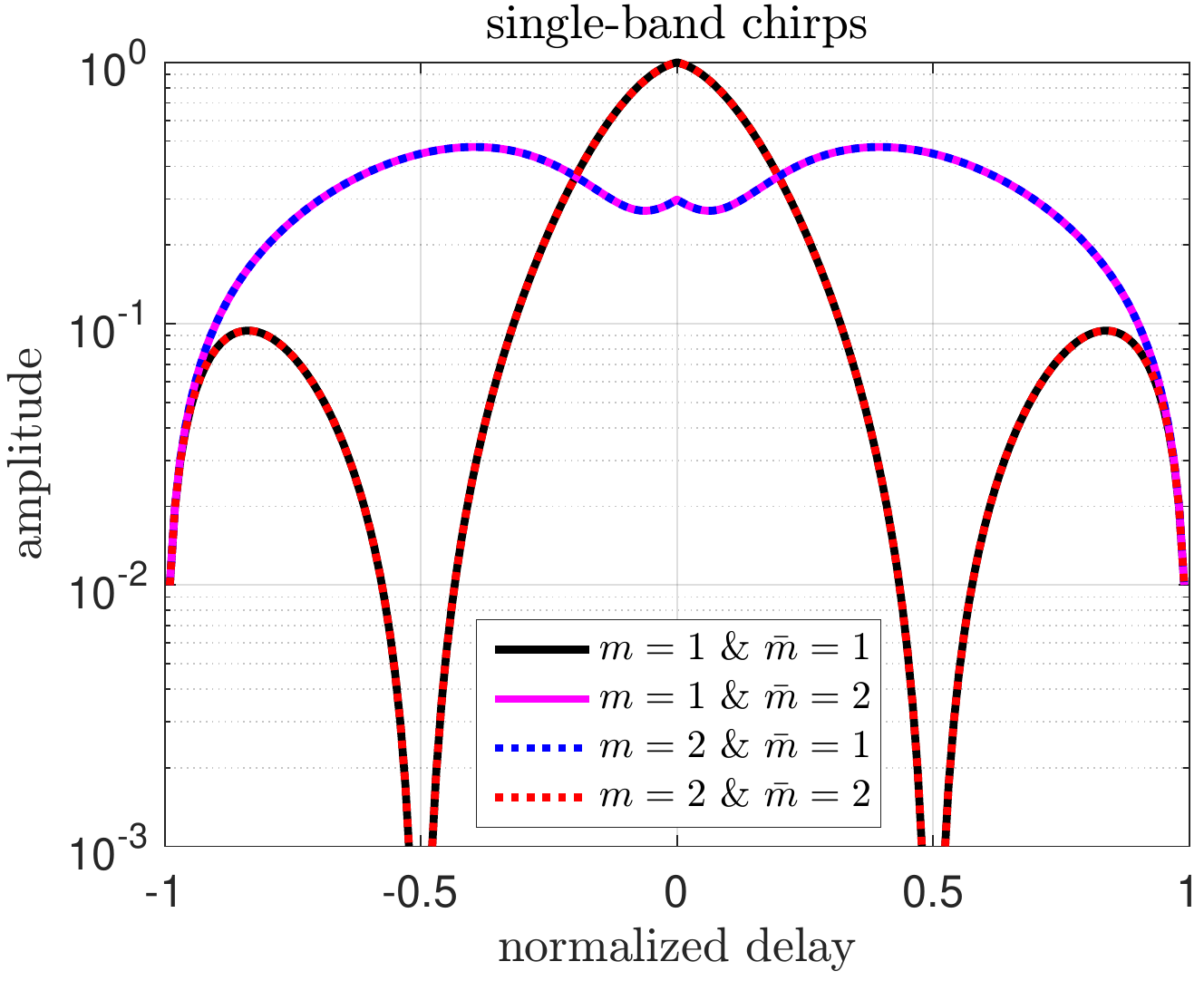}
\\
(b)
\et
\caption{Auto- and cross-ambiguity function versus the delay (normalized by the pulse duration $T_\text{p}$) with zero Doppler.}
\label{fig:AF}
\end{figure}
LFM waveforms, which are frequently used in radar, are employed as test waveforms. We consider two types of LFM waveforms with different ambiguity characteristics. The first are \emph{multi-band} chirps:
\begin{align}\label{eq:multibandchirp}
&p_{m}(t)=\frac{1}{\sqrt{T_\text{p}}}e^{\jmath\pi\beta( t^2/T_\text{p}+\eta m t)}\\
&0 \leq t \leq T_\text{p},\ m=1,\ldots,M,\notag
\end{align}
where $\beta$ is the bandwidth of the waveform and $\eta$ is a bandwidth gap parameter that is selected to keep the frequency bands of different waveforms non-overlapping. The ambiguity functions of the multi-band chirps can be obtained by using \eqref{eq:multibandchirp} in \eqref{equ:AF}, which are shown in Fig.\,\ref{fig:AF} (a) when $M=2$, $\eta=3$, and $\beta=400$ kHz. The multi-band chirps are \emph{orthogonal} with zero cross ambiguity at zero delay and Doppler, i.e., when the waveforms arrive at the RX \emph{synchronously}. However, they are not strictly orthogonal in distributed MIMO radar due to asynchronous propagation, but can be considered as \emph{approximately orthogonal} since the cross ambiguity is relatively low for small delay/Doppler offsets.

The second are \emph{single-band} chirps with overlapping instantaneous frequency. For $M=2$, we employ an up chirp given by
\ben\label{equ:up}
p_{\text{u}}(t)=\frac{1}{\sqrt{T_\text{p}}}e^{\jmath(\pi\beta t^2/T_\text{p}+\kappa\pi\beta t)},\,0\leq t<T_\text{p},
\een
and a down chirp
\ben\label{equ:down}
p_{\text{d}}(t)=\frac{1}{\sqrt{T_\text{p}}}e^{\jmath(-\pi\beta t^2/T_\text{p}+2\pi\beta t+\kappa\pi\beta t)},\,0\leq t\leq T_\text{p},
\een
where $\kappa$ is a constant that controls the center frequency of the chirps. The general expression of the single-band chirps can be found in \cite[eq.(9)]{WangLi20RadarConf}. Fig.\,\ref{fig:AF} (b) depicts the ambiguity functions of the single-band chirps when $M=2$, $\kappa=3$, and $\beta=400$ kHz, which shows the single-band chirps are \emph{non-orthogonal} waveforms with high cross ambiguity.
\subsection{Effect of Unequal Channel Strength}
\label{subsec:diverseresults}
\begin{figure}[t]
\centering
\bt{c}
\includegraphics[width=2.4in]{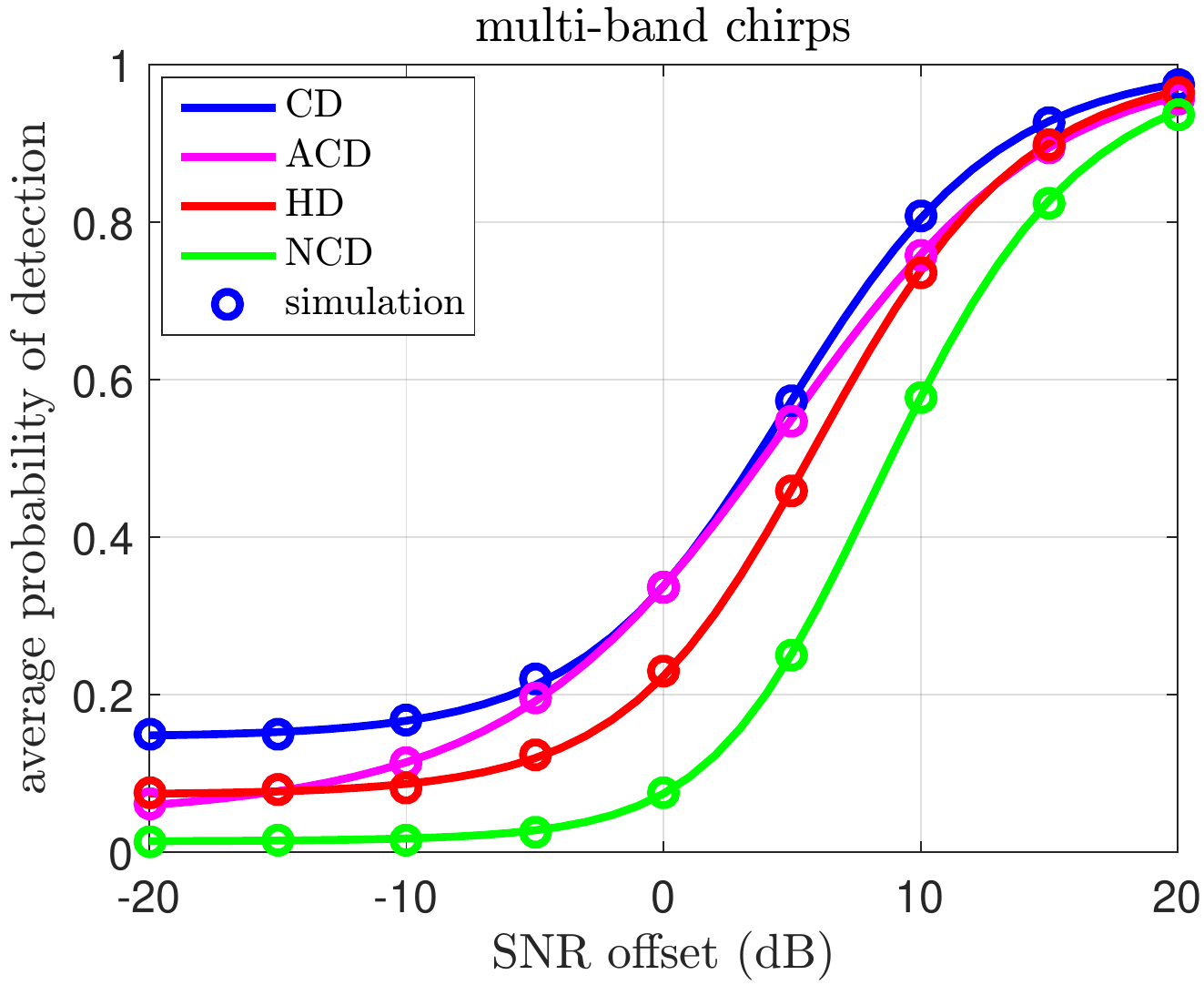}\\
(a)\\ 
\includegraphics[width=2.4in]{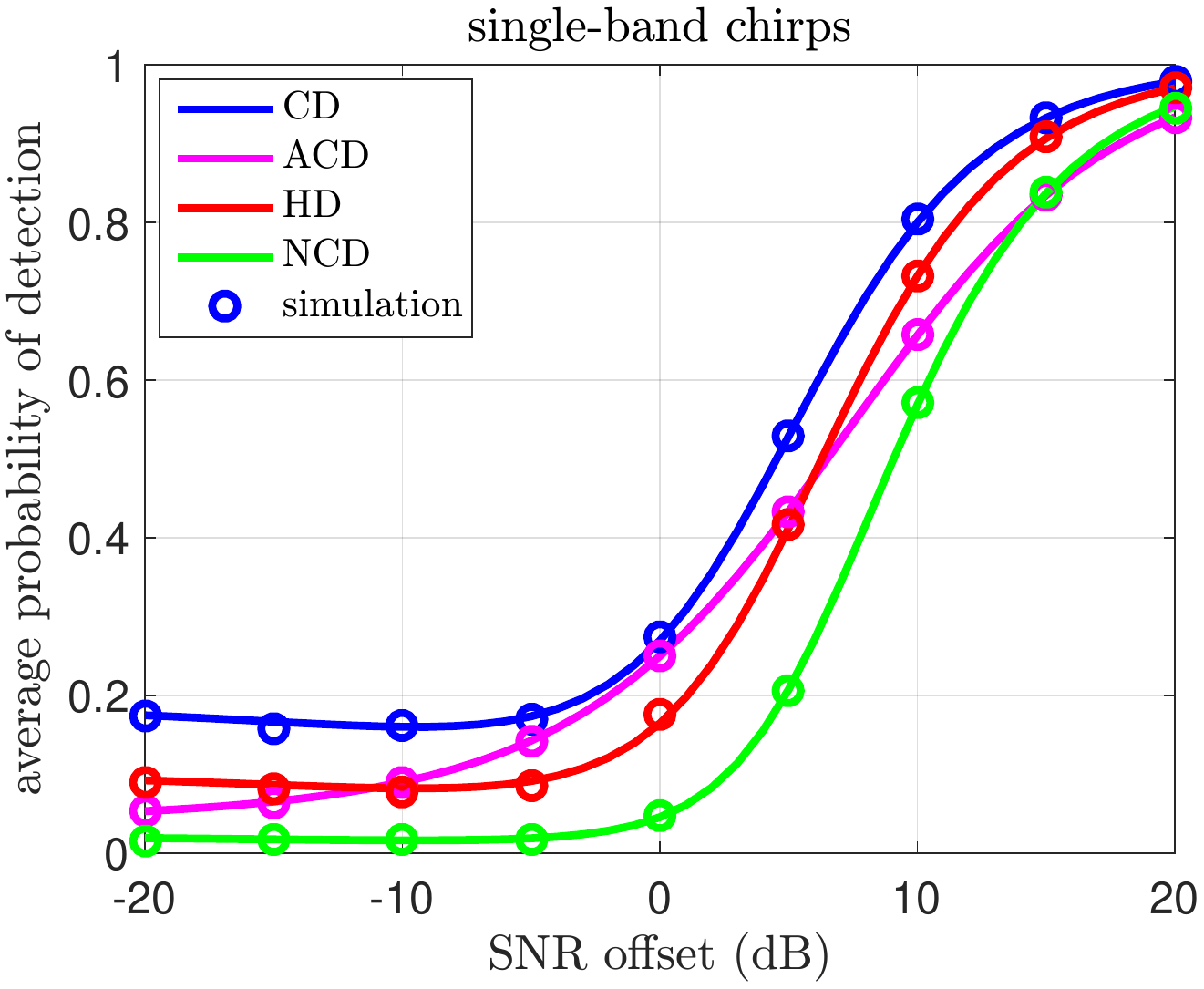}
\\
(b)
\et
\caption{$\bar{P}_d$ of distributed MIMO radar versus SNR offset $\text{SNR}_{21}-\text{SNR}_{11}$, where $\text{SNR}_{11}=0$ dB. The solid lines are obtained from theoretical analysis while the markers (circles) are obtained by simulation.}
\label{fig:weighting}
\end{figure}
We consider a scenario when the two propagation paths from the TXs to the RX have different SNR. In particular, we fix $\text{SNR}_{11}=0$ dB while varying $\text{SNR}_{21}$. Figs.\,\ref{fig:weighting} (a) and (b) depict the average probability of detection $\bar{P}_d$ versus $\text{SNR}_{21}$, where $\bar{P}_d$ is determined by using the theoretical analysis in Section \ref{sec:performanceanalysis} and simulation, respectively. It is seen that the analysis perfectly matches the computer simulation for all 4 detectors. With the multi-band chirps, Fig.\,\ref{fig:weighting} (a) shows that CD outperforms ACD when $\text{SNR}_{21}\neq\text{SNR}_{11}$, where the benefit comes from the amplitude weighting employed by CD. In addition, HD is slightly worse than ACD but outperforms NCD since it employs partial coherent combining within each CPI but non-coherent combining across different antennas.

With single-band chirps, the 4 detectors exhibit similar performance behaviors in Fig.\,\ref{fig:weighting} (b) except that the gap between CD and ACD is larger and, furthermore, CD outperforms ACD even at $\text{SNR}_{21}=\text{SNR}_{11}$. This is because with single-band chirps, the cross terms in the MF outputs \eqref{data cube} are more significant, which are accounted for by CD in its phase compensation but are neglected by ACD.
\subsection{Effect of Propagation-Induced Offsets and Cross Terms}
\begin{figure}[t]
\centering
\bt{c}
\includegraphics[width=2.4in]{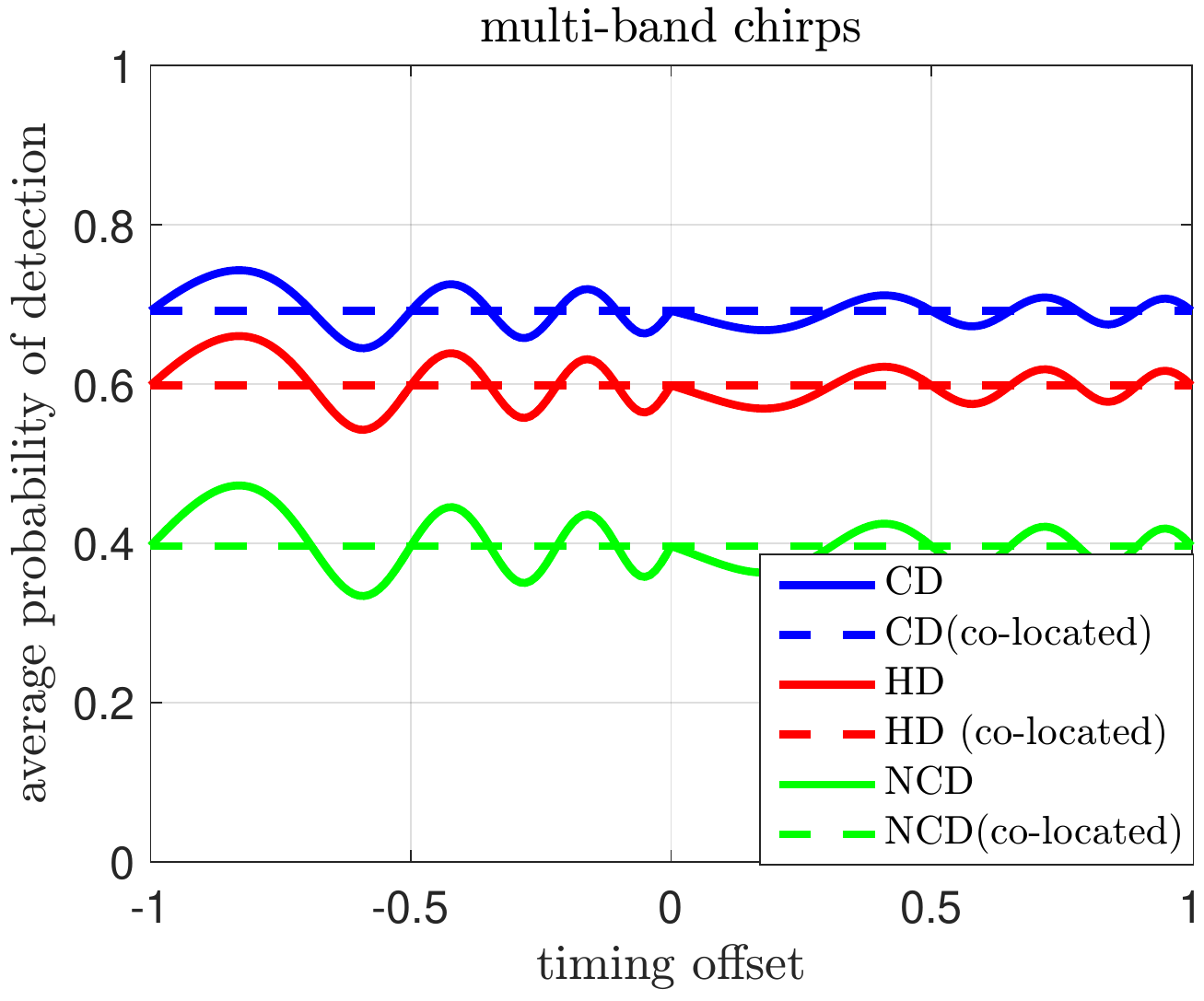}\\
(a)\\ 
\includegraphics[width=2.4in]{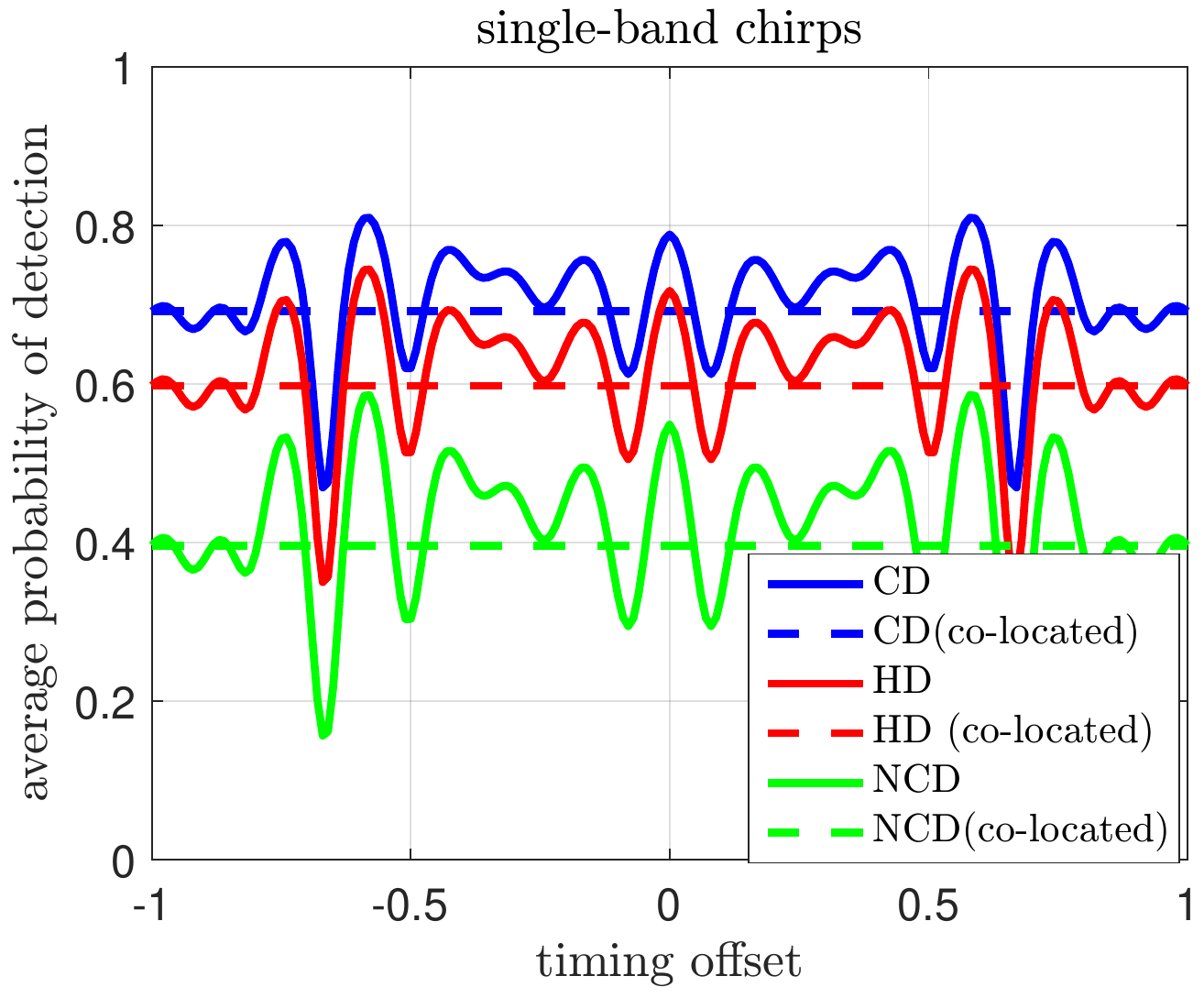}
\\
(b)
\et
\caption{$\bar{P}_d$ of distributed MIMO radar versus delay offset $\tau_{21}-\tau_{11}$ (normalized by the pulse duration $T_\text{p}$) in comparison with the synchronous co-located MIMO radar, when $\text{SNR}=0$ dB, $\psi_{11}=\psi_{21}=0$, and $f_{21}=f_{11}=0$.}
\label{fig:timingoffset}
\end{figure}
\begin{figure}[t]
\centering
\bt{c}
\includegraphics[width=2.4in]{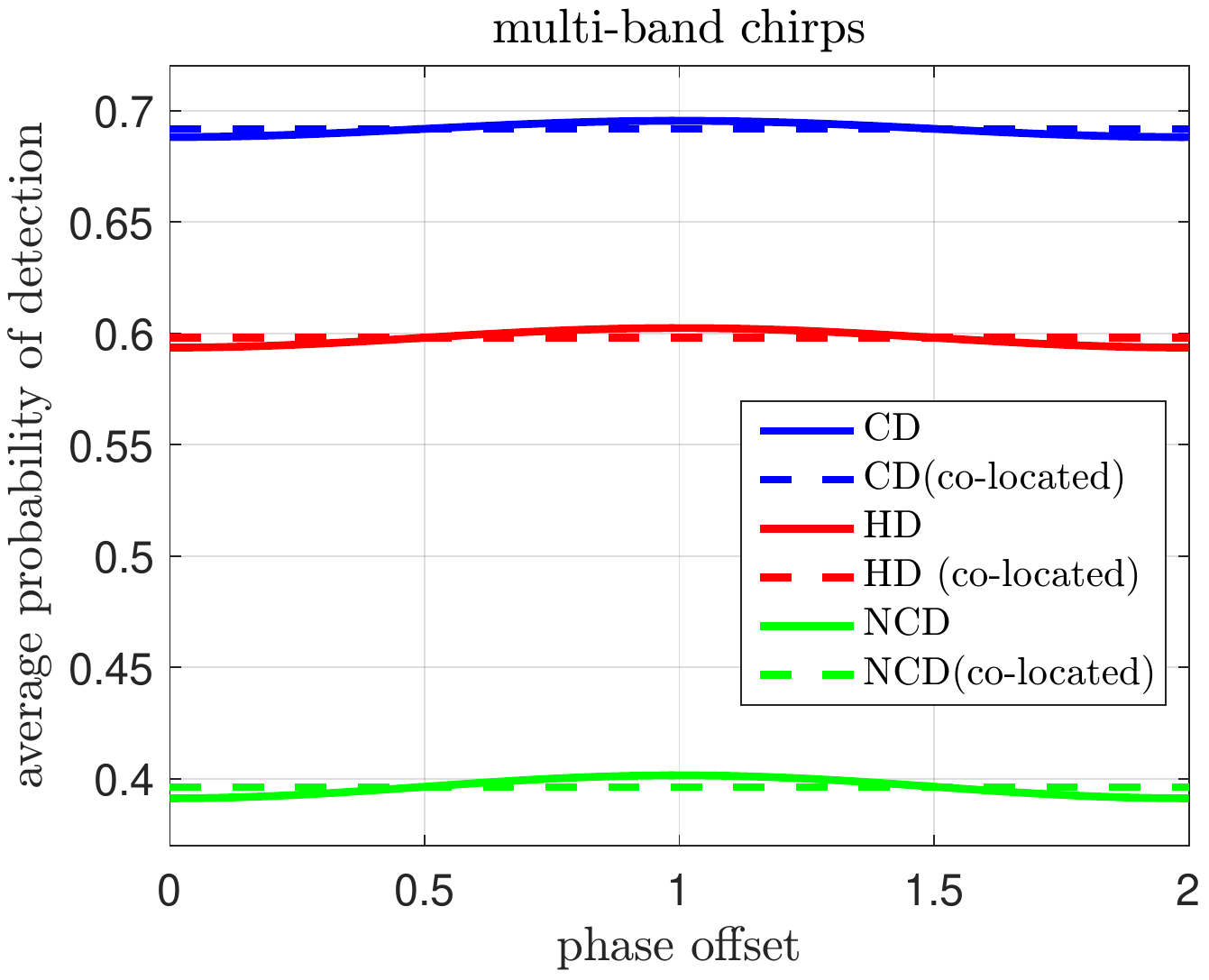}\\
(a)\\ 
\includegraphics[width=2.4in]{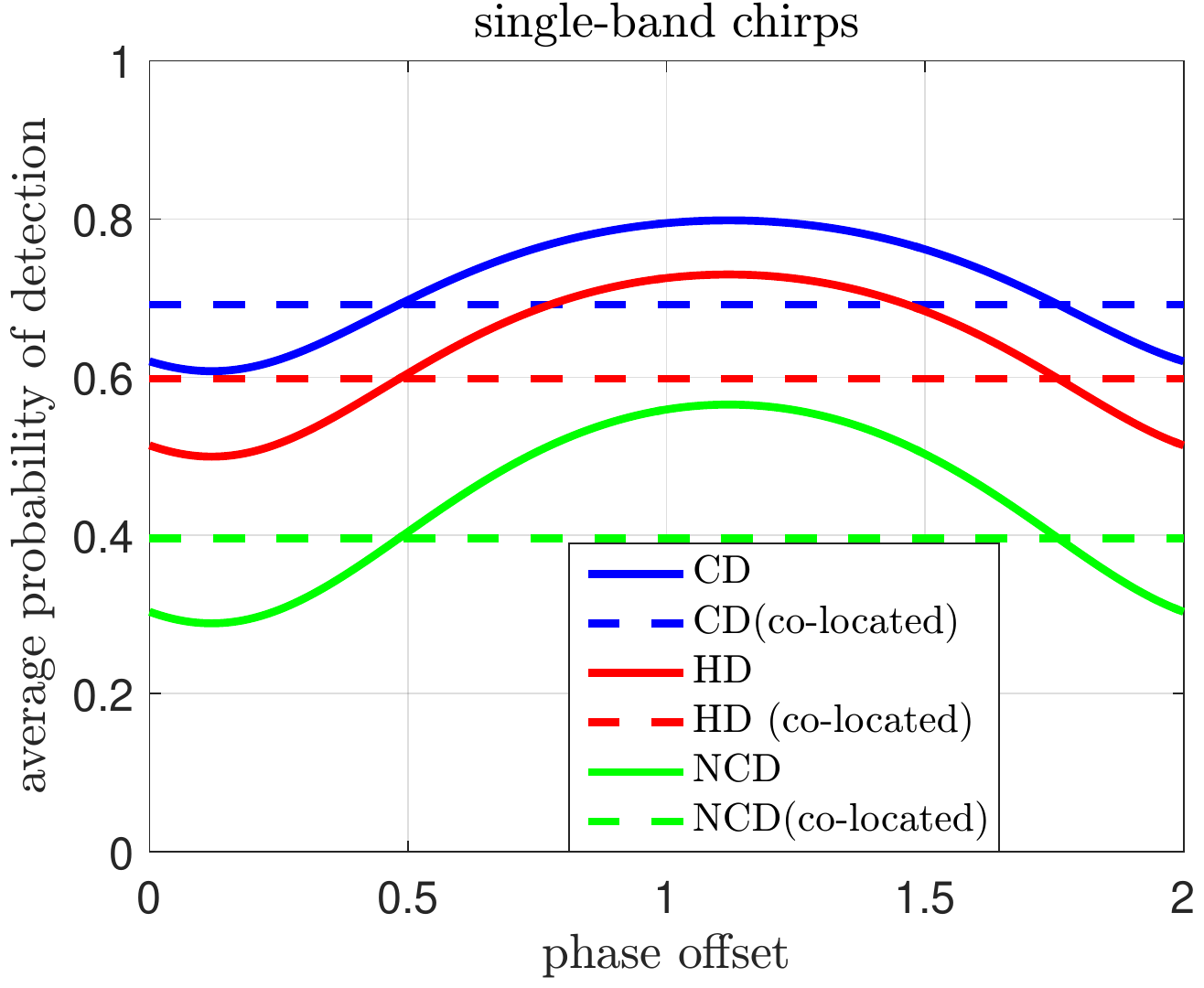}
\\
(b)
\et
\caption{$\bar{P}_d$ of distributed MIMO radar versus phase offset $\psi_{21}-\psi_{11}$ (normalized by $\pi$) in comparison with the synchronous co-located MIMO radar, when $\text{SNR}=0$ dB, $\tau_{11}=0.61T_\text{p}$, $\tau_{21}=0.1T_\text{p}$, and $f_{21}=f_{11}=0$.}
\label{fig:phaseoffset}
\end{figure}
\begin{figure}[t]
\centering
\bt{c}
\includegraphics[width=2.4in]{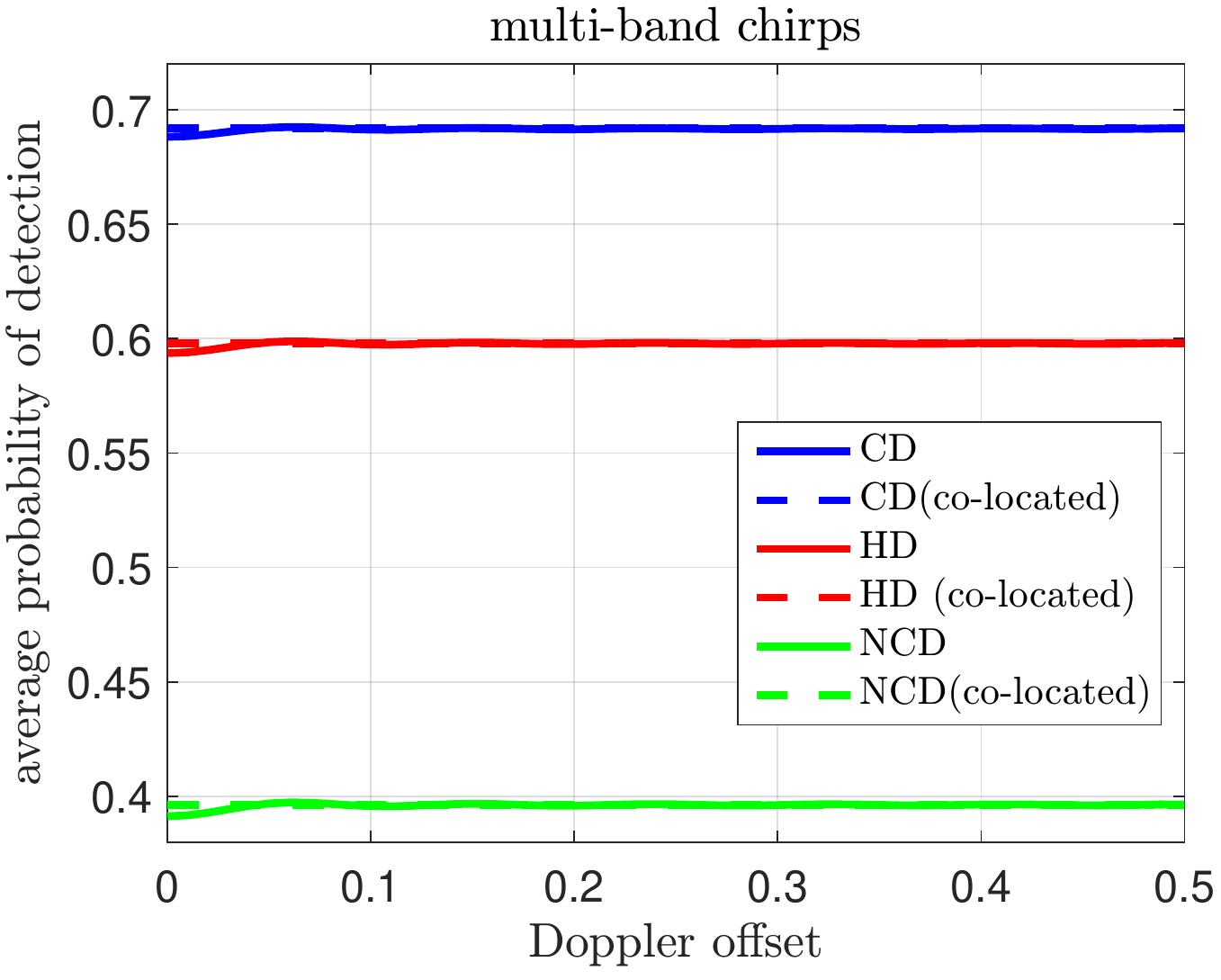}\\
(a)\\ 
\includegraphics[width=2.4in]{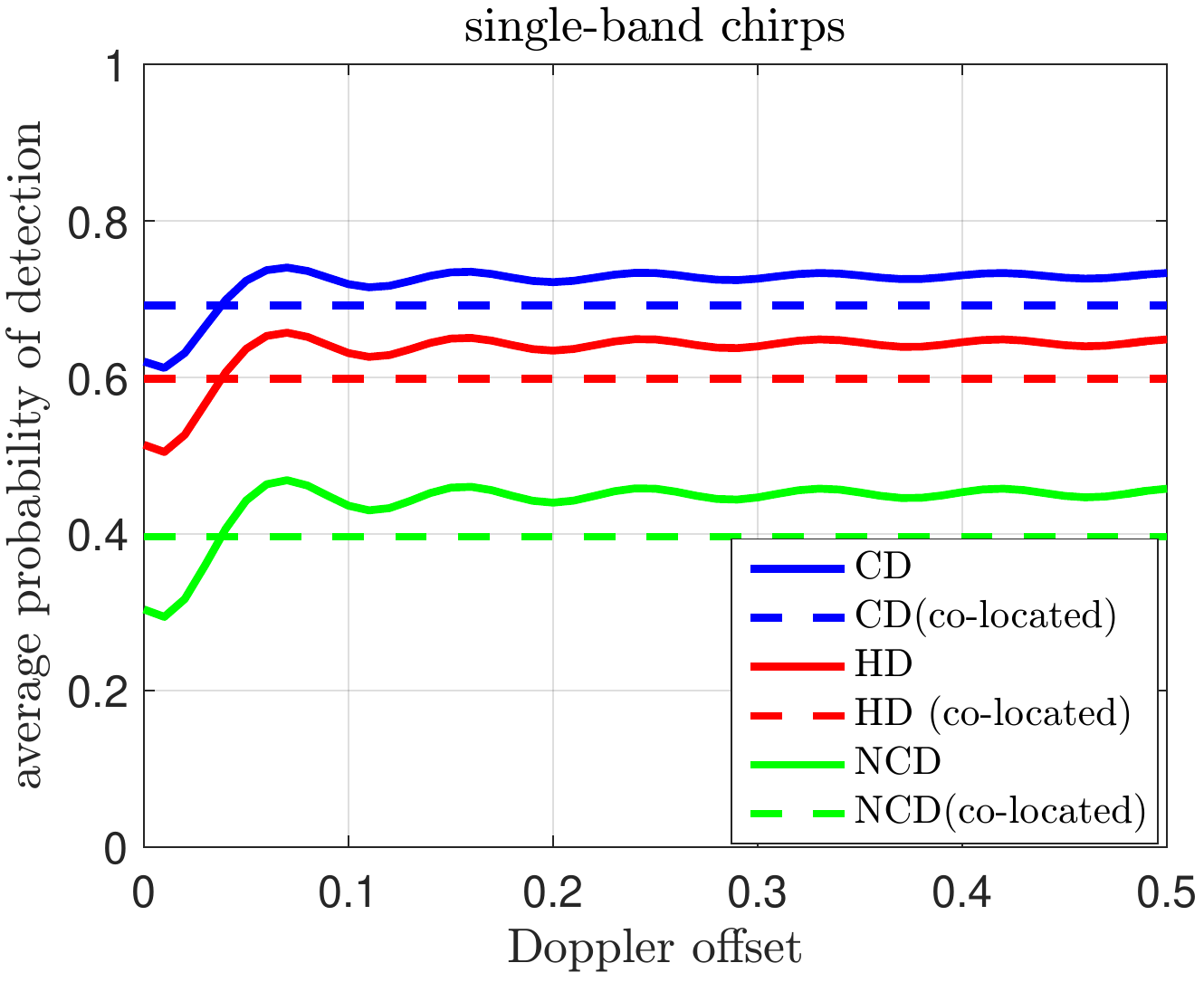}
\\
(b)
\et
\caption{$\bar{P}_d$ of distributed MIMO radar versus Doppler offset $f_{21}-f_{11}$ (normalized by the $\text{PRF}$) in comparison with the synchronous co-located MIMO radar, when $\text{SNR}=0$ dB, $\tau_{11}=0.61T_\text{p}$, $\tau_{21}=0.1T_\text{p}$, and $\psi_{11}=\psi_{21}=0$.}
\label{fig:Doppleroffset}
\end{figure}
As shown in Section \ref{sec:system_model_distribute}, asynchronous propagation is inherent in distributed MIMO radar, leading to inevitable offsets in delays, Doppler frequencies, and phases, as well as cross terms in the MF output \eqref{data cube}, even when the RX is perfectly synchronized with the TXs. Next, we examine the effect of such asynchronous propagation induced offsets on target detection. From now on, we no longer consider ACD, which is superseded by CD. In addition, we include the co-located MIMO radar as a benchmark, which assumes the TXs and RX are \emph{synchronous} with zero delay/Doppler/phase offsets and \emph{orthogonal} waveforms are employed, and as a result, there are \emph{no cross terms} in the MF output (see Section \ref{subsec:colocatedMIMO} and \ref{subsec:coMIMOdetection}).

Fig.\,\ref{fig:timingoffset} shows the performance of the CD, HD, and NCD of the distributed MIMO radar under various timing offsets, in comparison with the co-located MIMO radar, where $\bar{P}_d$ is computed analytically. It can be seen that the performance of all 3 detectors for distributed MIMO radar fluctuates as the delay offset varies. This is because the propagation delay affects the phase of the auto and cross terms, as shown in \eqref{data cube}. The auto and cross terms may add constructively when the difference of their phases is between $-\pi/2$ and $\pi/2$, or destructively when otherwise, which causes the fluctuation of the detection performance. A comparison between Figs.\,\ref{fig:timingoffset} (a) and (b) shows that the single-band chirps exhibit a larger fluctuation. This is because the single-band chirps have a larger cross terms than the multi-band chirps.

Figs.\,\ref{fig:phaseoffset} and \ref{fig:Doppleroffset} show the detection performance of the CD, HD, and NCD for the distributed MIMO radar versus phase offset and, respectively, Doppler offset, in comparison with the benchmark co-located MIMO radar. These detectors are seen to exhibit similar performance fluctuations as observed in Fig.\,\ref{fig:timingoffset} for similar reasons. Note that even though HD and NCD do not use phase information for detection, the observed signal varies with the phase offset, which leads to performance variation for these detectors.

It is interesting to note from Figs.\,\ref{fig:timingoffset} to \ref{fig:Doppleroffset} that the distributed MIMO radar, whose MF outputs include both auto- and cross-terms, may out- or under-perform the co-located MIMO radar, which only has the auto-terms, due to the aforementioned constructive or destructive addition. The challenge is that it is non-trivial to control how the auto- and cross-terms are added with each other, which are affected by many factors including the delays, phases, and Doppler frequencies of the propagation paths, as well as the ambiguities of the waveforms.
\subsection{Effect of Sync Errors}
\begin{figure}[t]
\centering
\bt{c}
\includegraphics[width=2.4in]{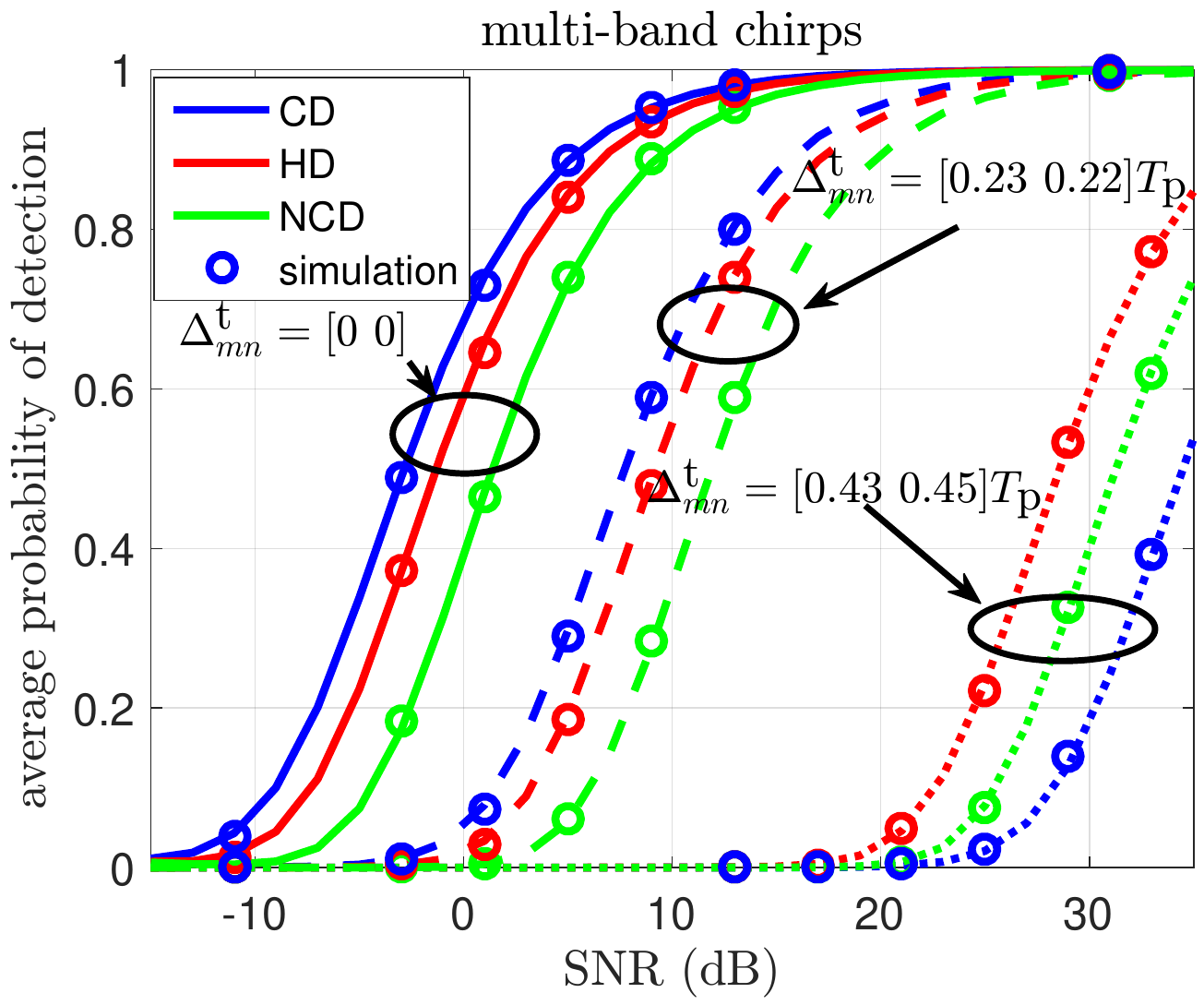}\\
(a)\\ 
\includegraphics[width=2.4in]{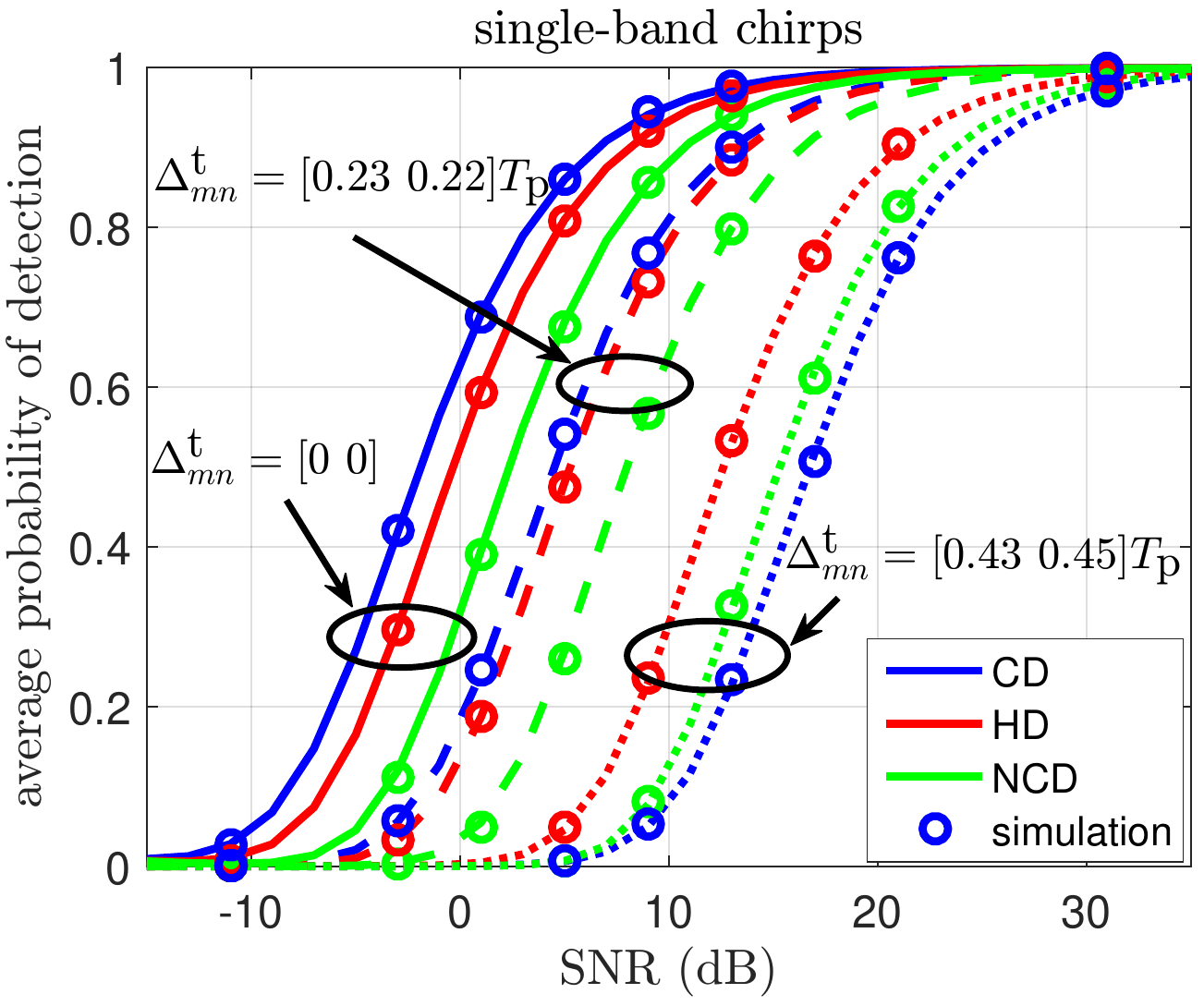}
\\
(b)
\et
\caption{$\bar{P}_d$ of distributed MIMO radar versus SNR without timing errors ($\Delta_{mn}^{\text{t}}=0$) or with two sets of timing errors. The solid lines are obtained from theoretical analysis while the markers (circles) are obtained by simulation.}
\label{fig:timingerror}
\end{figure}

\begin{figure}[t]
\centering
\bt{c}
\includegraphics[width=2.4in]{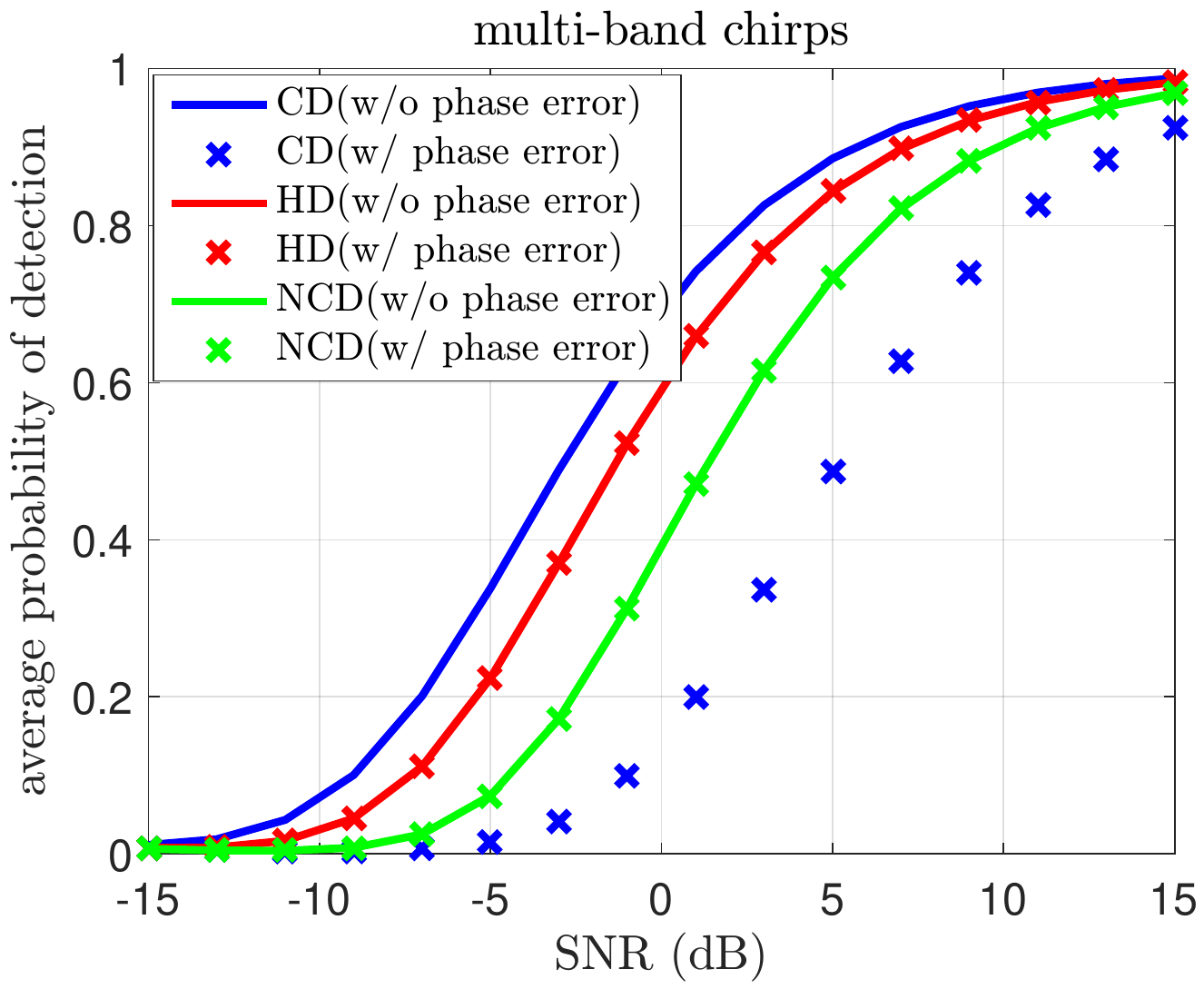}\\
(a)\\ 
\includegraphics[width=2.4in]{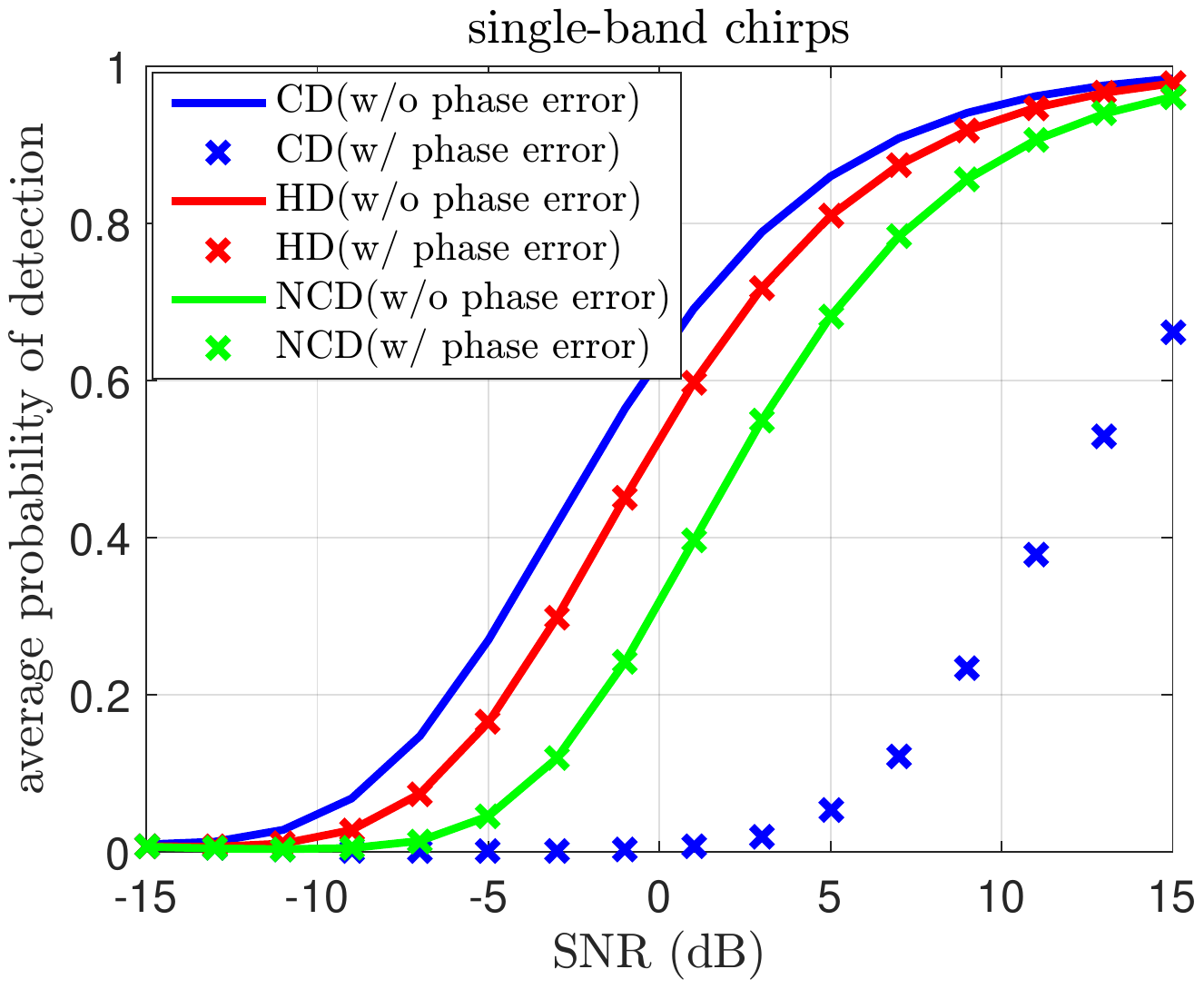}
\\
(b)
\et
\caption{$\bar{P}_d$ of distributed MIMO radar versus SNR without phase error ($\Delta_{mn}^{\text{p}}=0$) or with phase errors ($\Delta_{11}^{\text{p}}=0.053\pi$ and $\Delta_{21}^{\text{p}}=0.79\pi$).}
\label{fig:phaseerror}
\end{figure}
\begin{figure}[t]
\centering
\bt{c}
\includegraphics[width=2.4in]{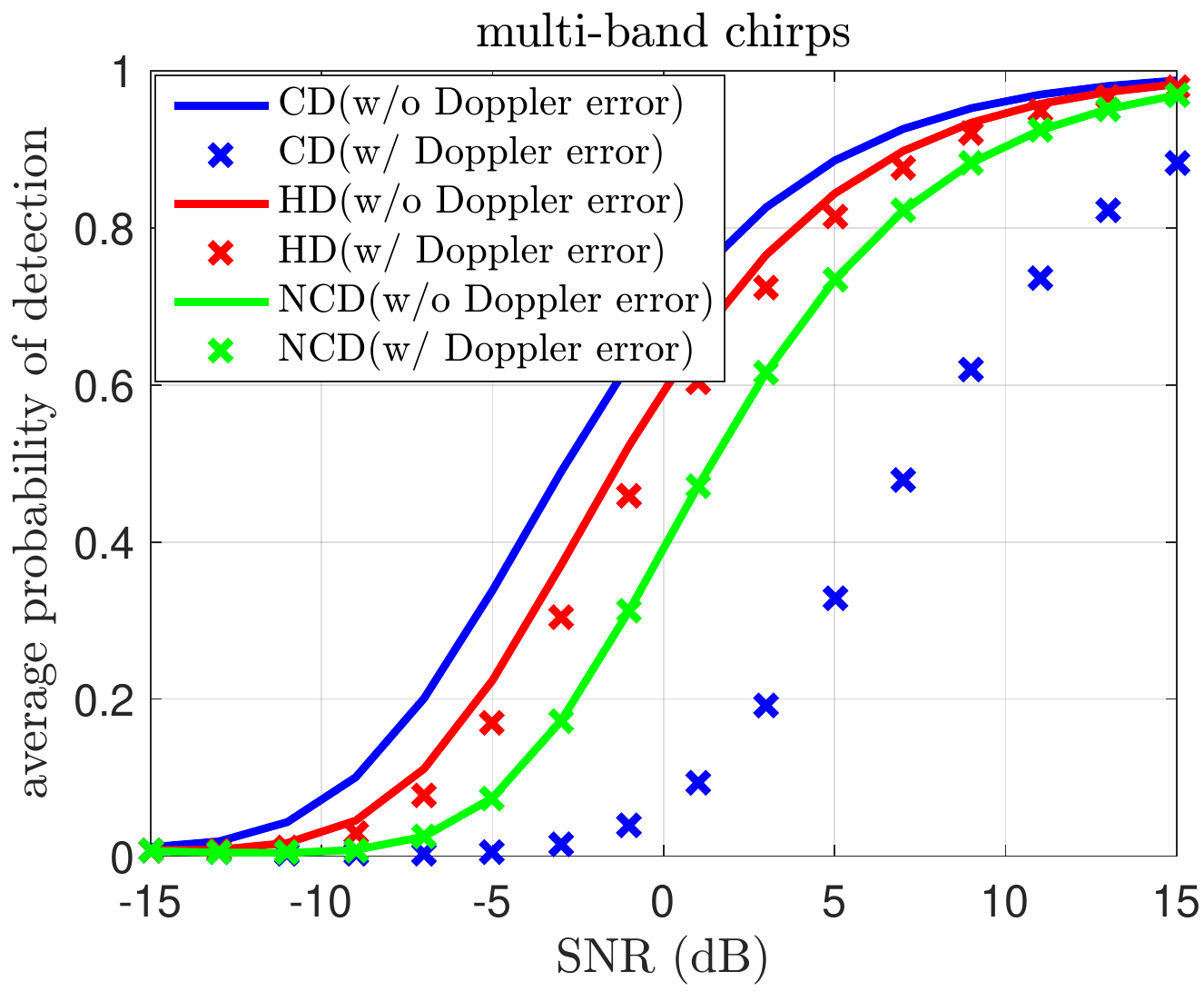}\\
(a)\\ 
\includegraphics[width=2.4in]{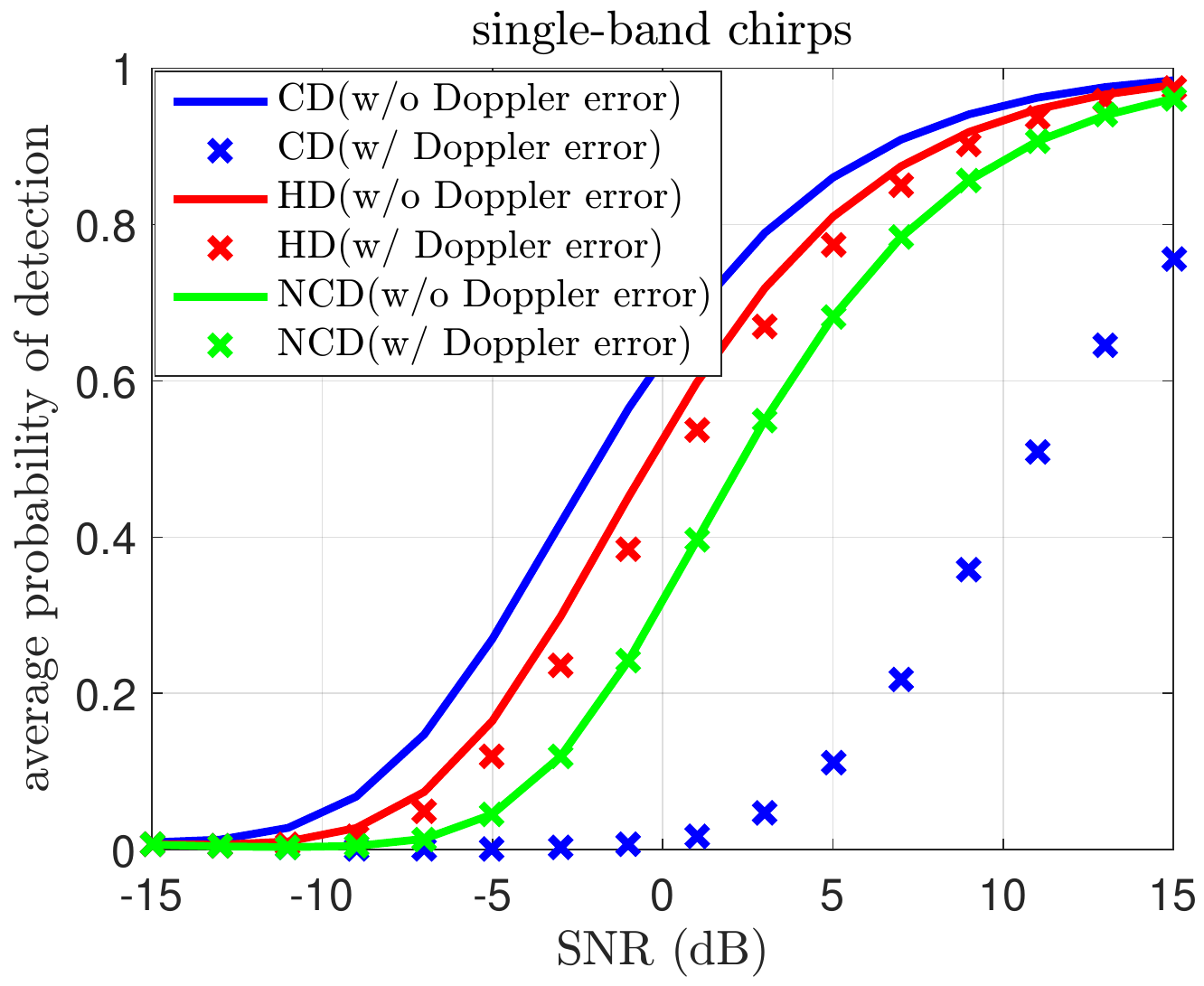}
\\
(b)
\et
\caption{$\bar{P}_d$ of distributed MIMO radar versus SNR without Doppler frequency errors ($\Delta_{mn}^{\text{f}}=0$ Hz) or with Doppler frequency errors ($\Delta_{11}^{\text{f}}=-25$ Hz and $\Delta_{21}^{\text{f}}=10$ Hz).}
\label{fig:Dopplererror}
\end{figure}
Finally, we evaluate the effects of sync errors, including timing,
phase, and Doppler frequency errors, on detection performance.
Fig.\,\ref{fig:timingerror} depicts the simulated and
analytical $\bar{P}_d$ for CD, HD, and NCD under various timing
conditions. Again, there is a perfect match between the simulated and
analytical results. In addition, it is seen that in general, as the
timing error increases, the performance of all 3 detectors
degrades. This is because a larger timing error implies the sampling
location is further away from the peak of the auto ambiguity function,
which results in a higher loss of the energy of the desired auto term
and the associated SNR. It was observed in \cite{WangLi20RadarConf},
if the timing error is much smaller than the reciprocal of the
waveform bandwidth but still significant relative to the carrier
period so that the SNR loss is negligible, then it will only impact
coherent detectors such as ACD as the timing-error-induced phase error
may not be negligible. The observation applies to CD as well. For
space limitation, we do not duplicate the result here.

The impact of phase error is shown in Fig.\,\ref{fig:phaseerror}. It
is observed that the phase error only affects CD, which is because the
implementation of HD and NCD does not require any knowledge of the
phase while the CD requires it for coherent integration across
antennas. On the other hand, Fig.\,\ref{fig:Dopplererror} shows the
impact of Doppler frequency error. It is seen that
  Doppler frequency error degrades the performance of both CD and HD
  but not that of NCD. This is because the Doppler knowledge is
    required for the implementation of CD and HD. Interestingly,
    HD outperforms NCD with or without the Doppler
    error, while it exhibits much better performance than CD when
    Doppler error is present.

\section{Conclusion}
\label{sec:conclusion}
We examined the impact of non-orthogonal waveforms and sync errors on
target detection in distributed MIMO radar. Our main contributions
include the general asynchronous signal model for distributed MIMO
radar, the new CD and HD detectors, a complete statistical analysis of
CD, HD, and the previously introduced NCD and ACD for distributed MIMO
radar with sync errors. Our results indicate that cross terms stemmed
from non-orthogonal waveforms can be beneficial or detrimental to
target detection, while sync errors in timing, frequency, and phase
have different impacts on different detectors. The fact that detection
can benefit from cross terms opens up future research possibilities
for TX-side encoding to reap such performance gain, if propagation
related delay/frequency offsets in delay can be made available to the
TXs. Another future topic of interest is to extend
the study to cases involving clutter and extended target detection.

\appendices
\section{Derivations of CD \eqref{eq:CD} and HD \eqref{eq:HD} }\label{appendexDetectors}
\subsection{Derivation of CD \eqref{eq:CD}}\label{appendexCD}
Coherent detection requires the knowledge of the Doppler steering matrices $\Sbf_n$, ambiguity function matrices $\Xcalbf_{mn}$, and channel vectors $\hbf_{mn}$ for phase and amplitude compensation. Given these estimates, a CD can be obtained by using a generalized likelihood ratio test (GLRT) framwork, which is detailed next.

Let $\Ybf\triangleq[\ybf_{11},\cdots,\ybf_{MN}]^T\in\Cset^{MN\times K}$, which contains observations from all RXs. According to \eqref{equ:hypothese}, the likelihood function under $\Hcal_1$ and $\Hcal_0$ can be expressed as
\begin{align}
&p_1(\Ybf;\alpha)=\frac{1}{(\pi\sigma^2)^{KMN}}\notag\\&\times\exp\Big(-\frac{1}{\sigma^2}\sum_{m=1}^M\sum_{n=1}^N\Vert\ybf_{mn}-\alpha\Sbf_n\Xcalbf_{mn}\hbf_{mn}\Vert^2\Big),\\
&p_0(\Ybf)=\frac{1}{(\pi\sigma^2)^{KMN}}\exp\Big(-\frac{1}{\sigma^2}\sum_{m=1}^M\sum_{n=1}^N\Vert\ybf_{mn}\Vert^2\Big).
\end{align}
It follows the log-likelihood ratio (LLR) is
\begin{align}
&l(\Ybf)=\log\frac{p_1(\Ybf;\alpha)}{p_0(\Ybf)}\notag\\
&=\frac{1}{\sigma^2}\sum_{m=1}^M\sum_{n=1}^N\Big(\Vert\ybf_{mn}\Vert^2-\Vert\ybf_{mn}-\alpha\Sbf_n\Xcalbf_{mn}\hbf_{mn}\Vert^2\Big).
\end{align}
The GLRT requires the maximum likelihood estimate (MLE) of $\alpha$ under $\Hcal_1$. Taking the derivative of the log-likelihood $\ln p_1(\Ybf;\alpha)$ w.r.t. $\alpha$ and setting it to zero yields the MLE
\begin{align}
\hat{\alpha}=\frac{\sum_{m=1}^M\sum_{n=1}^N(\Sbf_n\Xcalbf_{mn}\hbf_{mn})^H\ybf_{mn}}{\sum_{m=1}^M\sum_{n=1}^N(\Sbf_n\Xcalbf_{mn}\hbf_{mn})^H(\Sbf_n\Xcalbf_{mn}\hbf_{mn})}.
\end{align}
Substituting the MLE into the LLR:
\begin{align}
l(\Ybf)=\frac{\sum_{m=1}^M\sum_{n=1}^N\ybf_{mn}^H\Sbf_n(\Sbf_n^H\Sbf_n)^{-1}\Sbf_n^H\ybf_{mn}}{\sigma^2\sum_{m=1}^M\sum_{n=1}^N(\Sbf_n\Xcalbf_{mn}\hbf_{mn})^H(\Sbf_n\Xcalbf_{mn}\hbf_{mn})}.
\end{align}
The denominator of $l(\Ybf)$ can be absolved into the test threshold, which reduces the GLRT to
\begin{align}\label{eq:DCD}
\widetilde{T}_\text{CD}=\left\vert\sum_{m=1}^M\sum_{n=1}^N(\Sbf_n\Xcalbf_{mn}\hbf_{mn})^H\ybf_{mn}\right\vert^2.
\end{align}
For practical implementation, $\Sbf_n$, $\Xcalbf_{mn}$, and $\hbf_{mn}$ are constructed from delay/phase/Doppler estimates. In the presence of sync errors, they are formed by \Cref{eq:estimatedDoppler,equ:timingerrorAF,eq:estimatedphase}, in which case the CD is given by \eqref{eq:CD}.
\subsection{Derivation of HD \eqref{eq:HD}}\label{appendexHD}
The HD can be obtained by using GLRT and treating $\betabf_{mn}=\alpha\Xcalbf_{mn}\hbf_{mn}$, which lumps the target amplitude $\alpha$, ambiguity function matrix $\Xcalbf_{mn}$, and channel coefficient $\hbf_{mn}$, as an \emph{unstructured unknown vector}. Specifically, let $\betabf\triangleq[\betabf_{11},\cdots,\betabf_{MN}]^T\in\Cset^{MN\times K}$. According to \eqref{equ:hypothese}, the likelihood functions can be expressed as
\begin{align}
&p_1(\Ybf;\betabf)=\frac{1}{(\pi\sigma^2)^{KMN}}\notag\\&\times\exp\Big(-\frac{1}{\sigma^2}\sum_{m=1}^M\sum_{n=1}^N\Vert\ybf_{mn}-\Sbf_n\betabf_{mn}\Vert^2\Big),\\
&p_0(\Ybf)=\frac{1}{(\pi\sigma^2)^{KMN}}\exp\Big(-\frac{1}{\sigma^2}\sum_{m=1}^M\sum_{n=1}^N\Vert\ybf_{mn}\Vert^2\Big),
\end{align}
The MLE of $\betabf_{mn}$ is obtained by taking the derivative of $\ln p_1(\Ybf;\betabf)$ w.r.t. $\betabf_{mn}$ and setting it to zero:
\begin{align}
\hat{\betabf}_{mn}=(\Sbf_n^H\Sbf_n)^{-1}\Sbf_n^H\ybf_{mn}.
\end{align}
Using the above MLE in the LLR along with some simplifications leads to the following test statistic
\begin{align}
\widetilde{T}_\text{HD}&=\sum_{m=1}^M\sum_{n=1}^N\ybf_{mn}^H\Sbf_n(\Sbf_n^H\Sbf_n)^{-1}\Sbf_n^H\ybf_{mn}\notag\\
&=\sum_{m=1}^M\sum_{n=1}^N\Vert\Sbf_n(\Sbf_n^H\Sbf_n)^{-1}\Sbf_n^H\ybf_{mn}\Vert^2.
\end{align}
In practice, $\Sbf_n$ is replaced by $\hat{\Sbf}_n$ formed from \eqref{eq:estimatedDoppler} along with Doppler estimates, leading to \eqref{eq:HD}.
\section{Proofs of Theorems \ThmRef{theoremNCD} to \ThmRef{theoremHD}}\label{appendexAnalysis}
\subsection{Proof of Theorem \ThmRef{theoremNCD}}\label{appendexAnalysisNCD}
Consider the NCD \eqref{eq:NCD}, which is included below for easy reference (dropping the subscript ``NCD'' for simplicity)
\begin{align}
T= \sum_{m=1}^M\sum_{n=1}^N\ybf_{mn}^H\ybf_{mn}.
\end{align}
Clearly, $T$ is a square sum involving $2KMN$ independent and identically distributed (i.i.d.) real Gaussian random variables with variance $\sigma^2/2$ and zero mean (under $\Hcal_0$) and non-zero mean (under $\Hcal_1$), respectively. Hence,
\begin{equation}
T\backsim
\begin{cases}
&\frac{\sigma^2}{2}\mathcal{X}_{2KMN}^2\ \ \text{under}\ \mathcal{H}_{0}\\
&\frac{\sigma^2}{2}\mathcal{X}_{2KMN}^{'2}(\lambda)\ \ \text{under}\ \mathcal{H}_{1},
\end{cases}
\end{equation}
where $\mathcal{X}_{2KMN}^2$ and $\mathcal{X}_{2KMN}^{'2}(\lambda)$ denote the central and, respectively, noncentral chi-square distribution with $2KMN$ degrees of freedom and the noncentrality parameter $\lambda$ is given by \eqref{eq:lambdaNCD}. Based on the above distributions, the probability of false alarm and the probability of detection can be calculated as
\begin{align}\label{eq:PfNCD}
&P_f=\int_{\gamma}^{\infty}f(T\vert\mathcal{H}_0)dT\notag\\
&=\frac{2}{\sigma^2}\int_{\gamma}^{\infty}\frac{1}{2^{KMN}\Gamma(KMN)}\Big(\frac{2x}{\sigma^2}\Big)^{KMN-1}e^{-\frac{x}{\sigma^2}}dx\notag\\
&=\frac{\Gamma(KMN)-\bar{\Gamma}(KMN,\gamma/\sigma^2)}{\Gamma(KMN)}
\end{align}
\begin{align}\label{eq:PdNCD}
&P_d=\int_{\gamma}^{\infty}f(T\vert\mathcal{H}_1)dT\notag\\
&=\frac{1}{\sigma^2}\int_{\gamma}^{\infty}e^{-\frac{x+\lambda\sigma^2/2}{\sigma^2}}\Big(\frac{2x}{\lambda\sigma^2}\Big)^{\frac{KMN-1}{2}}I_{(KMN-1)}\Big(\sqrt{\frac{2\lambda x}{\sigma^2}}\Big)dx\notag\\
&=Q_{KMN}\Big(\sqrt{\lambda},\sqrt{\frac{2\gamma}{\sigma^2}}\Big).
\end{align}
\subsection{Proof of Theorem \ThmRef{theoremDCD}}\label{appendexAnalysisDCD}
Let $Y\triangleq \sum_{m=1}^M\sum_{n=1}^N (\hat{\Sbf}_n\hat{\Xcalbf}_{mn}\hat{\hbf}_{mn})^H\ybf_{mn}$. The test statistic of CD \eqref{eq:CD} can be written as
\begin{align}
T=\left\vert Y\right\vert^2.
\end{align}
It is easy to show that $Y$ is complex Gaussian with
\begin{gather}
\mathbb{E}\{Y\vert\mathcal{H}_0\}=0,\\ \mathbb{E}\{Y\vert\mathcal{H}_1\}=\alpha\sum_{m=1}^M\sum_{n=1}^N(\hat{\Sbf}_n\hat{\Xcalbf}_{mn}\hat{\hbf}_{mn})^H\Sbf_n\Xcalbf_{mn}\hbf_{mn},\\
\text{var}\{Y\vert\mathcal{H}_0\}=\text{var}\{Y\vert\mathcal{H}_1\}=\sigma^2\sum_{m=1}^M\sum_{n=1}^N\Vert\hat{\Sbf}_n\hat{\Xcalbf}_{mn}\hat{\hbf}_{mn}\Vert^2.
\end{gather}
Let
\begin{align}
\widetilde{T}\triangleq\frac{2T}{\sigma^2\sum_{m=1}^M\sum_{n=1}^N\Vert\hat{\Sbf}_n\hat{\Xcalbf}_{mn}\hat{\hbf}_{mn}\Vert^2}.
\end{align}
It follows that
\begin{equation}
\widetilde{T}\backsim
\begin{cases}
&\mathcal{X}_{2}^2\ \ \text{under}\ \mathcal{H}_{0}\\
&\mathcal{X}_{2}^{'2}(\lambda)\ \ \text{under}\ \mathcal{H}_{1},
\end{cases}
\end{equation}
where the noncentrality parameter $\lambda$ is given by \eqref{eq:lambdaDCD}. Hence, the probability of false alarm is given by
\begin{align}\label{eq:PfCD}
P_f=P(T>\gamma\vert\mathcal{H}_0)=P\Big(\widetilde{T}>\frac{2\gamma}{\varsigma\sigma^2}\Big\rvert\mathcal{H}_0\Big)=e^{-\frac{\gamma}{\varsigma\sigma^2}},
\end{align}
where $\varsigma$ is defined in \eqref{eq:varsigma}. Likewise, the probability of detection is given by
\begin{align}\label{eq:PdCD}
P_d&=P(T>\gamma\vert\mathcal{H}_1)=P(\widetilde{T}>\frac{2\gamma}{\varsigma\sigma^2}\vert\mathcal{H}_1)\notag\\
&=1-F_{\chi_2^{'2}(\lambda)}(\frac{2\gamma}{\varsigma\sigma^2})=Q_1(\sqrt{\lambda},\sqrt{\frac{2\gamma}{\varsigma\sigma^2}}),
\end{align}
where $F_{\chi_2^{'2}(\lambda)}(x)=1-Q_1(\sqrt{\lambda},\sqrt{x})$ is the cumulative distribution function (CDF) of the non-central chi-square random variable $\chi_2^{'2}(\lambda)$.
\subsection{Proof of Theorem \ThmRef{theoremACD}}\label{appendexAnalysisACD}
Define $X=\sum_{m=1}^M\sum_{n=1}^N\sum_{k=0}^{K-1} e^{-\jmath\hat{\theta}_{mnk}}y_{mn}(k)$. The test statistic of ACD \eqref{eq:teststatisticACD} is equivalent to
\begin{align}
T=\left\vert X\right\vert^2.
\end{align}
Based on the distribution of $\ybf_{mn}$, we can obtained the distribution of $y_{mn}(k)$ as $y_{mn}(k)\backsim\mathcal{CN}(0,\sigma^2)$ under $\Hcal_0$ and $y_{mn}(k)\backsim\mathcal{CN}(x_{mn}(k),\sigma^2)$ under $\Hcal_1$. Thus, we have
\begin{equation}
X\backsim
\begin{cases}
&\mathcal{CN}(0,\sigma^2)\ \ \text{under}\ \mathcal{H}_{0}\\
&\mathcal{CN}(\sum\limits_{m=1}^M\sum\limits_{n=1}^N\sum\limits_{k=0}^{K-1}e^{-\jmath\hat{\theta}_{mnk}}x_{mn}(k),\sigma^2)\ \ \text{under}\ \mathcal{H}_{1},
\end{cases}
\end{equation}
where $x_{mn}(k)$ is defined in \eqref{data cube}. The distribution of the test statistic becomes
\begin{equation}
T\backsim
\begin{cases}
&\frac{\sigma^2}{2}\mathcal{X}_{2}^2\ \ \text{under}\ \mathcal{H}_{0}\\
&\frac{\sigma^2}{2}\mathcal{X}_{2}^{'2}(\lambda)\ \ \text{under}\ \mathcal{H}_{1},
\end{cases}
\end{equation}
where the noncentrality parameter is given by \eqref{eq:lambdaHCD}. Then, the probability of false alarm is
\begin{align}\label{eq:PfACD}
P_f=P(T>\gamma\vert\mathcal{H}_0)=e^{-\frac{\gamma}{KMN\sigma^2}},
\end{align}
and the probability of detection is
\begin{align}\label{eq:PdACD}
P_d&=P(T>\gamma\vert\mathcal{H}_1)=1-F_{\chi_2^{'2}(\lambda)}(\frac{2\gamma}{KMN\sigma^2})\notag\\
&=Q_1(\sqrt{\lambda},\sqrt{\frac{2\gamma}{KMN\sigma^2}}).
\end{align}
\subsection{Proof of Theorem \ThmRef{theoremHD}}\label{appendexAnalysisHD}
The HD test statistic \eqref{eq:HD} can be written as
\begin{align}
T=\sum_{m=1}^M\sum_{n=1}^N\hat{\varphibf}_{mn}^H\hat{\Sbf}_n^H\hat{\Sbf}_n\hat{\varphibf}_{mn},
\end{align}
where $\hat{\varphibf}_{mn}=(\hat{\Sbf}_n^H\hat{\Sbf}_n)^{-1}\hat{\Sbf}_n^H\ybf_{mn}$, which is a complex Gaussian random vector since it is a linear transformation of $\ybf_{mn}$. Specifically, under $\mathcal{H}_1$, $\hat{\varphibf}_{mn}\backsim\mathcal{CN}(\varphibf_{mn},\Cbf_{\hat{\varphibf}_{mn}})$, where
\begin{gather}
\varphibf_{mn}=\alpha(\hat{\Sbf}_n^H\hat{\Sbf}_n)^{-1}\hat{\Sbf}_n^H\Sbf_n\Xcalbf_{mn}\hbf_{mn},\\
\Cbf_{\hat{\varphibf}_{mn}}=\sigma^2(\hat{\Sbf}_n^H\hat{\Sbf}_n)^{-1}.
\end{gather}
As a result, we have
\begin{align}
\frac{\hat{\varphibf}_{mn}^H\hat{\Sbf}_n^H\hat{\Sbf}_n\hat{\varphibf}_{mn}}{\sigma^2/2}=2\hat{\varphibf}_{mn}^H\Cbf_{\hat{\varphibf}_{mn}}^{-1}\hat{\varphibf}_{mn}.
\end{align}
Next, let $\varphibf_{mn}=\nubf_{mn}+\jmath\mubf_{mn}$ and $\hat{\varphibf}_{mn}=\hat{\nubf}_{mn}+\jmath\hat{\mubf}_{mn}$, and define the $2M\times1$ real vectors $\varepsilonbf_{mn}=[\nubf^T_{mn}\ \mubf^T_{mn}]^T$ and $\hat{\varepsilonbf}_{mn}=[\hat{\nubf}_{mn}^T\ \hat{\mubf}_{mn}^T]^T$. Then, the test statistic is equivalent to \cite[Appendix 15A]{Kay93estimation}:
\begin{align}
\frac{\hat{\varphibf}^H_{mn}\hat{\Sbf}_n^H\hat{\Sbf}_n\hat{\varphibf}_{mn}}{\sigma^2/2}=\hat{\varepsilonbf}_{mn}^T\Cbf_{\hat{\varepsilonbf}_{mn}}^{-1}\hat{\varepsilonbf}_{mn},
\end{align}
where $\Cbf_{\hat{\varepsilonbf}_{mn}}$ is the $2M\times 2M$ covariance matrix of the real vector $\hat{\varepsilonbf}_{mn}$. Since $\hat{\varphibf}_{mn}\backsim\mathcal{CN}(\varphibf_{mn},\Cbf_{\hat{\varphibf}_{mn}})$, we have
\begin{align}
\hat{\varepsilonbf}_{mn}\backsim\mathcal{N}(\varepsilonbf_{mn},\Cbf_{\hat{\varepsilonbf}_{mn}}).
\end{align}
Clearly, $\hat{\varepsilonbf}_{mn}^T\Cbf_{\hat{\varepsilonbf}_{mn}}^{-1}\hat{\varepsilonbf}_{mn}$ is central (under $\mathcal{H}_0$) and noncentral (under $\mathcal{H}_1$) chi-square distributed \cite[Section 2.3]{Kay98}. Equivalently, we have
\begin{equation}
\frac{\hat{\varphibf}_{mn}^H\hat{\Sbf}_n^H\hat{\Sbf}_n\hat{\varphibf}_{mn}}{\sigma^2/2}\backsim
\begin{cases}
&\mathcal{X}_{2M}^2\ \ \text{under}\ \mathcal{H}_{0}\\
&\mathcal{X}_{2M}^{'2}(\lambda_{mn})\ \ \text{under}\ \mathcal{H}_{1},
\end{cases}
\end{equation}
where
\begin{align}
&\lambda_{mn}=\varepsilonbf_{mn}^T\Cbf_{\hat{\varepsilonbf}_{mn}}^{-1}\varepsilonbf_{mn}=2\varphibf^H_{mn}\Cbf_{\hat{\varphibf}_{mn}}^{-1}\varphibf_{mn}\notag\\
&=\frac{\vert\alpha\vert^2\hbf_{mn}^H\Xcalbf_{mn}^H\Sbf_n^H\hat{\Sbf}_n(\hat{\Sbf}_n^H\hat{\Sbf}_n)^{-1}\hat{\Sbf}_n^H\Sbf_n\Xcalbf_{mn}\hbf_{mn}}{\sigma^2/2}.
\end{align}
According to the summation rule for a sum of weighted central/noncentral chi-square random variables \cite{Fernando05}, the original test statistic of the HD is given by
\begin{equation}
\frac{T}{\sigma^2/2}\backsim
\begin{cases}
&\mathcal{X}_{2NM^2}^2\ \ \text{under}\ \mathcal{H}_{0}\\
&\mathcal{X}_{2NM^2}^{'2}(\lambda)\ \ \text{under}\ \mathcal{H}_{1},
\end{cases}
\end{equation}
where $\lambda=\sum_{m=1}^M\sum_{n=1}^N\lambda_{mn}$. Based on the above distributions, the probability of false alarm and the probability of detection of the HD can be obtained by employing similar derivations as in \eqref{eq:PfNCD} and \eqref{eq:PdNCD}:
\begin{gather}\label{eq:PfHD}
P_f=\frac{\Gamma(NM^2)\bar{\Gamma}(NM^2,\gamma/\sigma^2)}{\Gamma(NM^2)}
\\
\label{eq:PdHD}
P_d=Q_{NM^2}\Big(\sqrt{\lambda},\sqrt{\frac{2\gamma}{\sigma^2}}\Big).
\end{gather}
\bibliographystyle{IEEEtran}
\bibliography{DMIMO}

\begin{thebibliography}{10}
\providecommand{\url}[1]{#1}
\csname url@samestyle\endcsname
\providecommand{\newblock}{\relax}
\providecommand{\bibinfo}[2]{#2}
\providecommand{\BIBentrySTDinterwordspacing}{\spaceskip=0pt\relax}
\providecommand{\BIBentryALTinterwordstretchfactor}{4}
\providecommand{\BIBentryALTinterwordspacing}{\spaceskip=\fontdimen2\font plus
\BIBentryALTinterwordstretchfactor\fontdimen3\font minus
  \fontdimen4\font\relax}
\providecommand{\BIBforeignlanguage}[2]{{%
\expandafter\ifx\csname l@#1\endcsname\relax
\typeout{** WARNING: IEEEtran.bst: No hyphenation pattern has been}%
\typeout{** loaded for the language `#1'. Using the pattern for}%
\typeout{** the default language instead.}%
\else
\language=\csname l@#1\endcsname
\fi
#2}}
\providecommand{\BIBdecl}{\relax}
\BIBdecl

\bibitem{LiStoica2007}
J.~{Li} and P.~{Stoica}, ``{MIMO} radar with colocated antennas,'' \emph{IEEE
  Signal Processing Magazine}, vol.~24, no.~5, pp. 106--114, 2007.

\bibitem{HaimovichBlum08}
A.~M. {Haimovich}, R.~S. {Blum}, and L.~J. {Cimini}, ``{MIMO} radar with widely
  separated antennas,'' \emph{IEEE Signal Processing Magazine}, vol.~25, no.~1,
  pp. 116--129, 2008.

\bibitem{WangLiHimedNon11}
P.~{Wang}, H.~{Li}, and B.~{Himed}, ``Moving target detection using distributed
  {MIMO} radar in clutter with nonhomogeneous power,'' \emph{IEEE Transactions
  on Signal Processing}, vol.~59, no.~10, pp. 4809--4820, 2011.

\bibitem{Govoni2015}
M.~A. Govoni, R.~Elwell, T.~Dogaru, and D.~Liao, ``A practical look at target
  detection using {MIMO} radar,'' in \emph{SPIE Defense, Security, and Sensing
  Conference}, vol. 9461, May 2015, pp. 228--238.

\bibitem{AubryMaioHuang16}
A.~{Aubry}, A.~{De Maio}, and Y.~{Huang}, ``{MIMO} radar beampattern design via
  {PSL}/{ISL} optimization,'' \emph{IEEE Transactions on Signal Processing},
  vol.~64, no.~15, pp. 3955--3967, 2016.

\bibitem{ChengHeLiaoFang18}
Z.~{Cheng}, Z.~{He}, B.~{Liao}, and M.~{Fang}, ``{MIMO} radar waveform design
  with {PAPR} and similarity constraints,'' \emph{IEEE Transactions on Signal
  Processing}, vol.~66, no.~4, pp. 968--981, 2018.

\bibitem{CuiMonga2020}
X.~{Yu}, K.~{Alhujaili}, G.~{Cui}, and V.~{Monga}, ``{MIMO} radar waveform
  design in the presence of multiple targets and practical constraints,''
  \emph{IEEE Transactions on Signal Processing}, vol.~68, pp. 1974--1989, 2020.

\bibitem{MaioVenturinoISIT08}
A.~{De Maio}, M.~{Lops}, and L.~{Venturino}, ``Diversity-integration trade-offs
  in {MIMO} detection,'' in \emph{2008 IEEE International Symposium on
  Information Theory}, 2008, pp. 594--598.

\bibitem{HeBlumTSP11}
Q.~{He} and R.~S. {Blum}, ``Diversity gain for {MIMO} {Neyman}-{Pearson} signal
  detection,'' \emph{IEEE Transactions on Signal Processing}, vol.~59, no.~3,
  pp. 869--881, 2011.

\bibitem{GodrichPoor2011}
H.~{Godrich}, A.~P. {Petropulu}, and H.~V. {Poor}, ``Power allocation
  strategies for target localization in distributed multiple-radar
  architectures,'' \emph{IEEE Transactions on Signal Processing}, vol.~59,
  no.~7, pp. 3226--3240, 2011.

\bibitem{HeBlumHaimovich10}
Q.~{He}, R.~S. {Blum}, H.~{Godrich}, and A.~M. {Haimovich}, ``Target velocity
  estimation and antenna placement for {MIMO} radar with widely separated
  antennas,'' \emph{IEEE Journal of Selected Topics in Signal Processing},
  vol.~4, no.~1, pp. 79--100, 2010.

\bibitem{LiangLeungSo2016}
J.~{Liang}, C.~S. {Leung}, and H.~C. {So}, ``Lagrange programming neural
  network approach for target localization in distributed {MIMO} radar,''
  \emph{IEEE Transactions on Signal Processing}, vol.~64, no.~6, pp.
  1574--1585, 2016.

\bibitem{HeLehmannBlum10}
Q.~{He}, N.~H. {Lehmann}, R.~S. {Blum}, and A.~M. {Haimovich}, ``{MIMO} radar
  moving target detection in homogeneous clutter,'' \emph{IEEE Transactions on
  Aerospace and Electronic Systems}, vol.~46, no.~3, pp. 1290--1301, 2010.

\bibitem{WangLiParametric13}
P.~{Wang}, H.~{Li}, and B.~{Himed}, ``A parametric moving target detector for
  distributed {MIMO} radar in non-homogeneous environment,'' \emph{IEEE
  Transactions on Signal Processing}, vol.~61, no.~9, pp. 2282--2294, May 2013.

\bibitem{YiBlum2020}
W.~{Yi}, T.~{Zhou}, Y.~{Ai}, and R.~S. {Blum}, ``Suboptimal low complexity
  joint multi-target detection and localization for non-coherent {MIMO} radar
  with widely separated antennas,'' \emph{IEEE Transactions on Signal
  Processing}, vol.~68, pp. 901--916, 2020.

\bibitem{AbramovichFrazer08}
Y.~I. {Abramovich} and G.~J. {Frazer}, ``Bounds on the volume and height
  distributions for the {MIMO} radar ambiguity function,'' \emph{IEEE Signal
  Processing Letters}, vol.~15, pp. 505--508, 2008.

\bibitem{AkcakayaNehorai11}
M.~{Akcakaya} and A.~{Nehorai}, ``{MIMO} radar sensitivity analysis for target
  detection,'' \emph{IEEE Transactions on Signal Processing}, vol.~59, no.~7,
  pp. 3241--3250, 2011.

\bibitem{WangLi2013Radarconf}
P.~{Wang}, H.~{Li}, and B.~{Himed}, ``Moving target detection for distributed
  {MIMO} radar with imperfect waveform separation,'' in \emph{2013 IEEE Radar
  Conference (RadarCon13)}, 2013, pp. 1--5.

\bibitem{WangLi2020}
P.~{Wang} and H.~{Li}, ``Target detection with imperfect waveform separation in
  distributed {MIMO} radar,'' \emph{IEEE Transactions on Signal Processing},
  vol.~68, pp. 793--807, 2020.

\bibitem{SunTangJunTSP14}
P.~{Sun}, J.~{Tang}, S.~{Wan}, and N.~{Zhang}, ``Identifiability analysis of
  local oscillator phase self-calibration based on hybrid
  {Cram$\acute{\text{e}}$r}-{Rao Bound} in {MIMO} radar,'' \emph{IEEE
  Transactions on Signal Processing}, vol.~62, no.~22, pp. 6016--6031, 2014.

\bibitem{YangBlum11}
Y.~{Yang} and R.~S. {Blum}, ``Phase synchronization for coherent {MIMO} radar:
  Algorithms and their analysis,'' \emph{IEEE Transactions on Signal
  Processing}, vol.~59, no.~11, pp. 5538--5557, 2011.

\bibitem{AkcakayaNehoraiPhase10}
M.~{Akçakaya} and A.~{Nehorai}, ``{MIMO} radar detection and adaptive design
  under a phase synchronization mismatch,'' \emph{IEEE Transactions on Signal
  Processing}, vol.~58, no.~10, pp. 4994--5005, 2010.

\bibitem{HeBlumCRB10}
Q.~{He} and R.~S. {Blum}, ``{Cram$\acute{\text{e}}$r}-{Rao Bound} for {MIMO}
  radar target localization with phase errors,'' \emph{IEEE Signal Processing
  Letters}, vol.~17, no.~1, pp. 83--86, Jan 2010.

\bibitem{DingChen2015}
L.~{Ding} and W.~{Chen}, ``{MIMO} radar sparse imaging with phase mismatch,''
  \emph{IEEE Geoscience and Remote Sensing Letters}, vol.~12, no.~4, pp.
  816--820, 2015.

\bibitem{ZhangVorobyov2016}
W.~{Zhang} and S.~A. {Vorobyov}, ``Joint robust transmit/receive adaptive
  beamforming for {MIMO} radar using probability-constrained optimization,''
  \emph{IEEE Signal Processing Letters}, vol.~23, no.~1, pp. 112--116, 2016.

\bibitem{HuLiuLiu2016}
Q.~{Hu}, H.~{Su}, S.~{Zhou}, Z.~{Liu}, and J.~{Liu}, ``Target detection in
  distributed {MIMO} radar with registration errors,'' \emph{IEEE Transactions
  on Aerospace and Electronic Systems}, vol.~52, no.~1, pp. 438--450, 2016.

\bibitem{WangLi20RadarConf}
F.~{Wang}, C.~{Zeng}, H.~{Li}, and M.~A. {Govoni}, ``Detection performance of
  distributed {MIMO} radar with asynchronous propagation and timing/phase
  errors,'' in \emph{2020 IEEE International Radar Conference (RADAR)}, April
  2020, pp. 13--18.

\bibitem{Richards05}
M.~A. Richards, \emph{Fundamentals of Radar Signal Processing}.\hskip 1em plus
  0.5em minus 0.4em\relax New York, NY, USA: McGraw-Hill, 2005.

\bibitem{prudnikov1986integrals}
A.~Prudnikov, Y.~Brychkov, I.~Brychkov, and O.~Marichev, \emph{Integrals and
  Series: More special functions}.\hskip 1em plus 0.5em minus 0.4em\relax
  Gordon and Breach Science Publishers, 1986.

\bibitem{Sofotasios2014}
P.~C. {Sofotasios}, M.~{Valkama}, T.~A. {Tsiftsis}, Y.~A. {Brychkov},
  S.~{Freear}, and G.~K. {Karagiannidis}, ``Analytic solutions to a {Marcum}
  {Q}-function-based integral and application in energy detection of unknown
  signals over multipath fading channels,'' in \emph{2014 9th International
  Conference on Cognitive Radio Oriented Wireless Networks and Communications
  (CROWNCOM)}, 2014, pp. 260--265.

\bibitem{Kay93estimation}
S.~M. Kay, \emph{Fundamentals of Statistical Signal Processing: Estimation
  Theory}.\hskip 1em plus 0.5em minus 0.4em\relax Upper Saddle River, NJ:
  Prentice Hall, 1993.

\bibitem{Kay98}
------, \emph{Fundamentals of Statistical Signal Processing: Detection
  Theory}.\hskip 1em plus 0.5em minus 0.4em\relax Upper Saddle River, NJ:
  Prentice Hall, 1998.

\bibitem{Fernando05}
L.-B. Fernando and A.~Castaño-Martínez, ``Distribution of a sum of weighted
  noncentral chi-square variables,'' \emph{TEST: An Official Journal of the
  Spanish Society of Statistics and Operations Research}, vol.~14, pp.
  397--415, 02 2005.

\end{thebibliography}
\end{document}